\documentclass[journal=jctccd,manuscript=article]{achemso}
\setcitestyle{super,open={},close={}}

\usepackage{hyperref, graphicx, xspace,subcaption}
\hypersetup{hidelinks, allcolors=black, colorlinks=true}
\usepackage{ragged2e}
\usepackage{multirow,booktabs,siunitx,natmove,threeparttable}
\usepackage{amsmath,amssymb}
\usepackage[version=4]{mhchem}
\usepackage{cleveref}
\usepackage{physics, calrsfs, mathtools, bm}
\usepackage{soul}
\usepackage{xcolor}
\usepackage[utf8]{inputenc}
\usepackage{comment}
\DeclareUnicodeCharacter{2212}{\textminus}

\author{Aniruddha Seal}
\affiliation{Department of Chemistry and Chicago Center for Theoretical Chemistry, University of Chicago, Chicago, IL 60637, USA}

\author{Bhavnesh Jangid}
\affiliation{Department of Chemistry and Chicago Center for Theoretical Chemistry, University of Chicago, Chicago, IL 60637, USA}

\author{Helen S. Clifford}
\affiliation{Department of Chemistry and Chicago Center for Theoretical Chemistry, University of Chicago, Chicago, IL 60637, USA}

\author{Andrew L. Ferguson}
\affiliation{Department of Chemistry and Chicago Center for Theoretical Chemistry, University of Chicago, Chicago, IL 60637, USA}
\altaffiliation{Pritzker School of Molecular Engineering, University of Chicago, Chicago, IL 60637, USA}

\author{Laura Gagliardi}
\affiliation{Department of Chemistry and Chicago Center for Theoretical Chemistry, University of Chicago, Chicago, IL 60637, USA}
\altaffiliation{Pritzker School of Molecular Engineering, University of Chicago, Chicago, IL 60637, USA}
\email{lgagliardi@uchicago.edu}

\title{Machine Learning the Complete Active-Space Wavefunction Reconstruction}

\begin{tocentry}
  \centering
  \includegraphics[width=6.67cm,height=12.23cm,keepaspectratio]{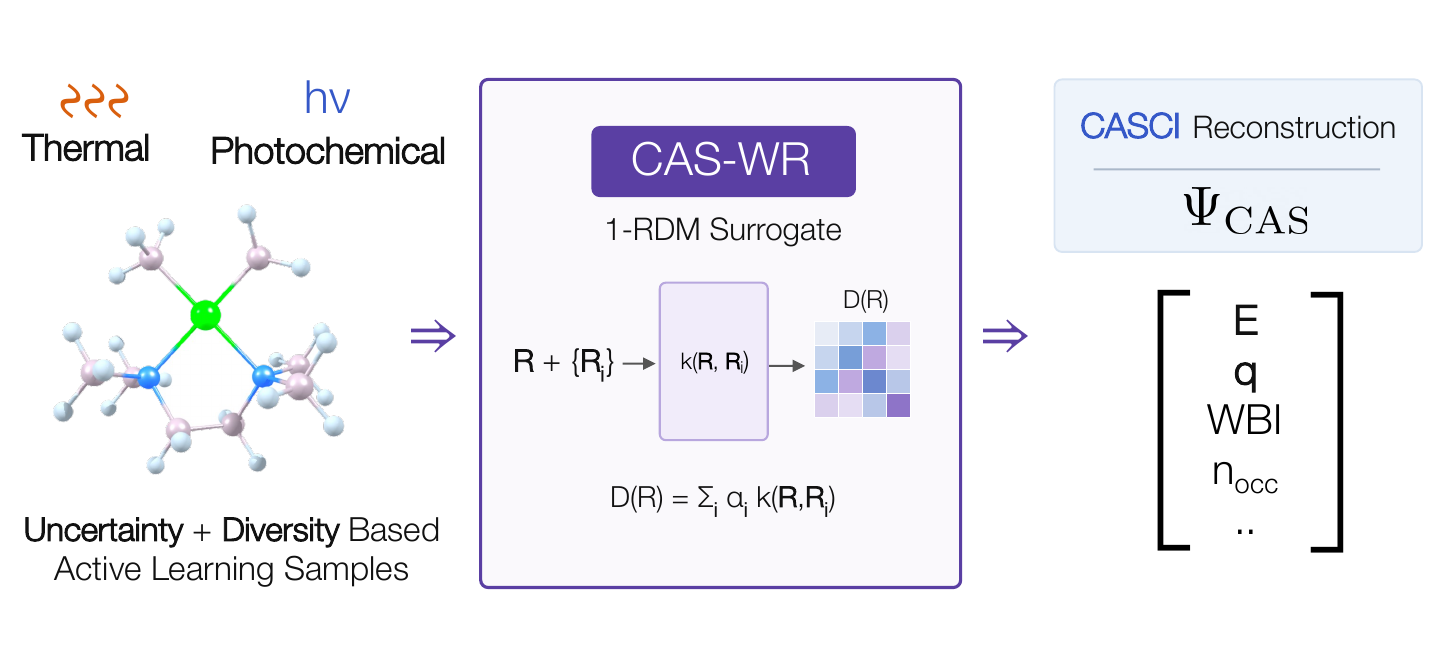}
\end{tocentry}

\begin{document}

\begin{abstract}
\noindent 

Machine-learned interatomic potentials enable multireference-quality reaction dynamics but do not retain the electronic structure underlying the predicted energies and forces. Recovering this information by independently performing complete active space self-consistent field (CASSCF) calculations is both expensive and potentially inconsistent. We introduce the Complete Active-Space Wave Function Reconstruction (CAS-WR) model, which learns the geometry-dependent electronic structure from CASSCF data and reconstructs the CAS wave function without repeated orbital optimization. Combined with uncertainty-based active learning, CAS-WR is demonstrated for the Diels--Alder reaction and \ce{TiC+}-catalyzed C--H activation. The approach enables trajectory-resolved analysis of multiconfigurational electronic structure across reactive ensembles. Preliminary applications to excited-state proton transfer and transition-metal photocatalysis suggest that CAS-WR can provide a general route toward recovering electronic structure along complex ground- and excited-state reactive dynamics.
\end{abstract}

\section{Introduction}

Machine-learned interatomic potentials (MLIPs) have transformed molecular simulation by enabling trajectories that approach the accuracy of an underlying electronic-structure method at a small fraction of its computational cost.\cite{deringer_machine_2019,schran_machine_2021,unke_machine_2021} Their accuracy has made it possible to study chemical reactions through long trajectories and statistically meaningful ensembles of reactive events. However, conventional MLIPs can only learn the potential energy and its nuclear derivatives. The electronic degrees of freedom are integrated out, 
and the resulting simulations do not retain direct access to the electronic structure associated with each molecular geometry.
Consequently, a MLIP trajectory cannot by itself reveal how the electronic structure evolves during bond breaking and formation.

This limitation restricts the use of MLIPs for electronic structure-informed design. Quantities such as orbital occupations, configuration weights, charge and spin redistribution, and changes in bond order often provide the most direct description of a reaction mechanism. They are also needed to identify orbital-based principles for modifying catalysts, chromophores, and reactive intermediates.\cite{cui_photocatalyst_2025,wang_eg_2019,zhang_spin_2024,suntivich_design_2011} In the absence of an explicit electronic representation, such information must be recovered through additional electronic-structure calculations on sub-selected structures,\cite{yan_atomistic_2026,yang_reactant-induced_2023} yet it does not provide the ensemble-level description.

A growing class of machine-learning methods addresses this limitation by learning electronic-structure quantities rather than only energies and forces.\cite{tang_deep-learning_2025,hindges_recent_2026} These approaches have targeted effective Hamiltonians and wave functions,\cite{schutt_unifying_2019,cignoni_electronic_2024,gong_general_2023} electron densities,\cite{grisafi_transferable_2018} and reduced density matrices,\cite{shao_machine_2023} thereby retaining access to a broader range of electronic observables than conventional MLIPs. Most such methods, however, have focused on mean-field methods like Hartree--Fock, Kohn--Sham density functional theory, or other predominantly single-reference descriptions, making their extension to reactions with substantial static correlation nontrivial. In regimes such as bond dissociation, open-shell intermediates, avoided crossings, and transition-metal-mediated reactions, multiple electronic configurations may contribute with comparable weight, motivating explicitly multiconfigurational descriptions.\cite{wardzala_multireference_2026} Neural quantum Monte Carlo methods provide an alternative by representing correlated wave functions directly with neural-network ans{\"a}tze and have shown high accuracy for strongly correlated systems and excited states.\cite{pfau_ab_2020,hermann_deep-neural-network_2020,pfau_accurate_2024,foster_ab_2026} These approaches, nevertheless, require variational optimization, making independent wave function optimization across the many configurations encountered in a reaction-dynamics ensemble computationally demanding.

Complete active space self-consistent field (CASSCF) theory provides a natural framework for describing the multiconfigurational electronic structure of reactive systems.\cite{roos_complete_1980,roos_complete_1987} For CASSCF, both the configuration-interaction coefficients in a selected active space and the molecular orbitals defining that active space are variationally optimized. In our previous work, the weighted active space protocol (WASP) enabled the generation of consistent active-spaces across diverse geometries and the construction of MLIPs with multireference (MC-PDFT\cite{li_manni_multiconfiguration_2014,sharma_multiconfiguration_2021}) accuracy.\cite{seal_weighted_2025} 
This approach made multireference-quality reaction dynamics practical, but the resulting MLIPs still supplied only energies and forces. The CASSCF wave function and the electronic observables derived from it were not retained.

Reconstructing this information by independently repeating a CASSCF calculation at every geometry is problematic for reasons beyond computational cost. CASSCF orbital optimization is a nonlinear, non-convex optimization problem with multiple local minima, and independent calculations at neighboring trajectory geometries may therefore converge to different stationary solutions.\cite{saade_excited_2024,marie_excited_2023} This is particularly problematic for reaction trajectories, where substantial structural changes can alter the orbital landscape and cause the optimization to switch between different CASSCF solutions. Repeating CASSCF calculations \textit{post-hoc} at selected trajectory geometries does not guarantee recovery of the same adiabatic surface, potentially leading to discontinuous or inconsistent electronic-structure assignments between the sampled geometries. The relevant goal is therefore not merely to obtain a converged CASSCF solution at each geometry, but to consistently recover the adiabatic surface represented by the multireference MLIP.

For the active space sizes that one can afford in reaction dynamics, a substantial fraction of the CASSCF computational cost arises from iterative orbital optimization. If suitable active-space orbitals can instead be supplied directly, the CI problem can be solved by CASCI without further orbital optimization. Approximate orbitals for CASCI can be obtained, for example, from unrestricted natural orbitals derived from broken-symmetry Hartree--Fock or from MP2 natural orbitals.\cite{bofill_unrestricted_1989,abrams_natural_2004} These approaches can reduce the cost relative to CASSCF, but still require an iterative mean-field or post-mean-field calculation at each new geometry and do not explicitly preserve the multireference electronic structure associated with the trajectory. A complementary strategy was introduced by Booth and co-workers, where they used eigenvector continuation to interpolate correlated many-body wave functions by solving a generalized eigenvalue problem within a model space spanned by a set of reference wave functions.\cite{rath_interpolating_2025} These developments provide useful alternatives to repeated CASSCF orbital optimization and illustrate different routes to recovering active-space orbitals or correlated wave functions. In the context of multireference MLIP trajectories, it is also desirable to recover the corresponding electronic structure efficiently, consistently with the underlying adiabatic surface, and in a form that retains access to wave function derived observables.

Here, we introduce the complete active space wave function reconstruction (CAS-WR) model, a machine-learning framework that reconstructs the CAS wave function without repeated orbital optimization. CAS-WR learns a geometry-dependent representation of the converged CASSCF solution from reference data generated across configuration space. From this representation, the molecular orbitals are recovered and used in a subsequent CASCI calculation. The approach therefore does not merely predict selected electronic properties. It reconstructs the multiconfigurational wave function, from which configuration weights, natural-orbital occupations, charge and spin distributions, bond-order measures, and other wavefunction-derived observables can be evaluated.

Combined with uncertainty-based active learning,\cite{seal_weighted_2025,perego_data_2024} we demonstrate CAS-WR for the Diels--Alder reaction of 1,4-butadiene and ethylene and the \ce{TiC+}-catalyzed C--H activation of methane. The resulting framework enables ensemble analysis of charge and spin redistribution, bond-order measures, configuration weights, natural-orbital occupations, and other properties derived from the reconstructed wave function. Therefore, this extends multireference-accurate MLIP dynamics from nuclear motion to trajectory-resolved electronic-structure analysis. Finally, we provide a preliminary demonstration of CAS-WR for excited-state transition-metal photocatalysis, highlighting its potential extension to excited states.

The central challenge is to identify a representation that is compact, varies smoothly with nuclear geometry, and retains sufficient information to recover the reference CASSCF solution. We consider two representations: the one-electron reduced density matrix $\mathbf D$ and the effective Fock matrix\cite{roos_complete_1980} $\mathbf F$ of the converged CASSCF solution,
\begin{equation}
\mathbf{D}_{pq}
=
\bra{
\Psi}
\hat a_p^\dagger \hat a_q
\ket{\Psi},
\end{equation}
and
\begin{equation}
\mathbf{F}_{pq}
=
h_{pq}
+
\sum_{rs}
\mathbf{D}_{rs}
\left[
(pq|rs)
-
\frac{1}{2}(pr|qs)
\right].
\end{equation}

\begin{figure}[h!]
    \centering
    \includegraphics[width=\linewidth]{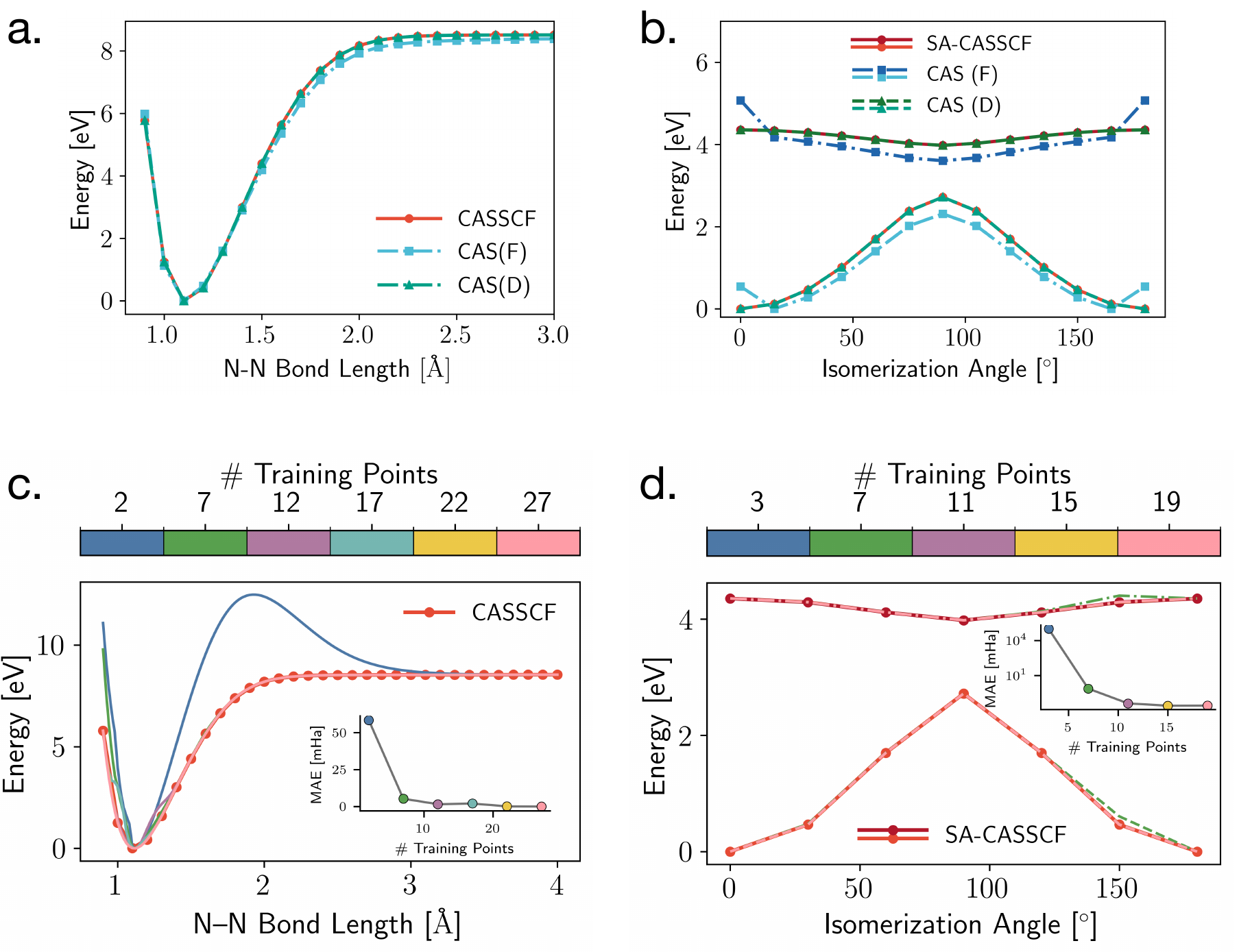}
    \caption{Assessment of electronic representations for CAS-WR. (a) Ground-state N$_2$ dissociation profiles from reference CASSCF and CASCI calculations using CAS(6e,6o)/cc-pVDZ. The CASCI orbitals are obtained by diagonalizing either the CASSCF effective Fock matrix, CAS($\mathbf F$), or the 1-RDM, CAS($\mathbf D$). (b) Corresponding excited-state double-bond torsion profiles for fulvene using SA(2)-CAS(6e,6o)/cc-pVDZ. Orbitals obtained from $\mathbf D$ reproduce the reference profiles more closely, particularly for fulvene. (c) CAS-WR reconstruction of the N$_2$ dissociation profile with increasing numbers of training geometries. (d) Corresponding CAS-WR reconstruction of the fulvene torsion profile.}
    \label{fig:1}
\end{figure}

 We assess their ability to reconstruct ground- and excited-state CAS wave functions for processes involving strong electron correlation, namely the ground-state dissociation of \ce{N2} and the excited-state isomerization of fulvene.\cite{cramer_essentials_2013,bearpark_can_1996} Natural orbitals were obtained by solving $(\mathbf S\mathbf D\mathbf S)\mathbf C_D=\mathbf S\mathbf C_D\mathbf n$, while Fock-derived orbitals were obtained from $\mathbf F\mathbf C_F=\mathbf S\mathbf C_F\boldsymbol{\varepsilon}$, where S is the AO overlap matrix; each orbital set was used in CASCI, and the resulting energies were compared with the reference CASSCF energies. \Cref{fig:1} shows that orbitals reconstructed from $\mathbf D$ consistently yield better agreement with the reference CASSCF results, with a particularly pronounced improvement for the excited-state isomerization of fulvene. This advantage can be understood from the information encoded by the two representations: the 1-RDM directly captures the correlated one-electron structure through its natural orbitals and their occupations, whereas $\mathbf F$ is an effective one-electron quantity whose eigenvectors reflect the multiconfigurational wave function only indirectly. This limitation can become especially pronounced in state-averaged calculations, where constructing $\mathbf F$ from weighted-averaged densities further reduces its state-specific orbital character. This interpretation is supported by the orbital overlaps with the reference CASSCF solution, as discussed in Section~SI of the Supporting Information. The poorer recovery obtained from $\mathbf F$ also explains why Marquetand and co-workers required subsequent orbital optimization when reconstructing wave functions from machine-learned Fock matrices.\cite{van_workum_casnet_2023} Because such optimization would introduce substantial overhead in trajectory-wide electronic-structure analyses, we chose $\mathbf D$ as the target quantity for CAS-WR.

The 1-RDM provides a compact representation intrinsically connected to configuration mixing and static correlation.\cite{zhang_multiconfigurational_2020} Having established the CASSCF 1-RDM as the target representation, we next require a geometry-dependent electronic descriptor from which the 1-RDM can be predicted. We use the one-electron Hamiltonian $\mathbf h(\mathbf R)$, which provides an operator-level representation of the nuclear geometry and has recently been explored as an input for learning electronic-structure quantities.\cite{nigam_machine_2026,shao_machine_2023} Following the work of Pavanello and co-workers,\cite{shao_machine_2023,rana_learning_2025} we formulate CAS-WR using kernel ridge regression (KRR),\cite{hastie_elements_2009} which provides a nonlinear but data-efficient interpolation of the electronic structure over the sampled region of configuration space. We write the predicted CASSCF 1-RDM as
\begin{equation}
\widehat{\mathbf D}_{\text{CAS-WR}}\!\left[\mathbf h(\mathbf R)\right]
=
\sum_{i=1}^{N_{\mathrm{train}}}
\boldsymbol{\beta}_i
K\!\left(
\mathbf h_i,
\mathbf h(\mathbf R)
\right),
\end{equation}
where $K$ is the kernel function, and $\boldsymbol{\beta}_i$ contains the learned regression coefficients for the elements of $\mathbf D$. We first evaluated CAS-WR on the \ce{N2} dissociation and fulvene isomerization benchmarks by reconstructing their state-averaged (SA)-CASSCF potential-energy profiles, as shown in \Cref{fig:1}. Accurate recovery required training geometries not only at the extrema of the reaction coordinate, but also at intermediate configurations that capture the evolution of the electronic structure between them. This indicates that extending CAS-WR to multidimensional systems requires an automated strategy for identifying and selecting the geometries most informative for training.

\begin{figure}[h!]
    \centering
    \includegraphics[width=\linewidth]{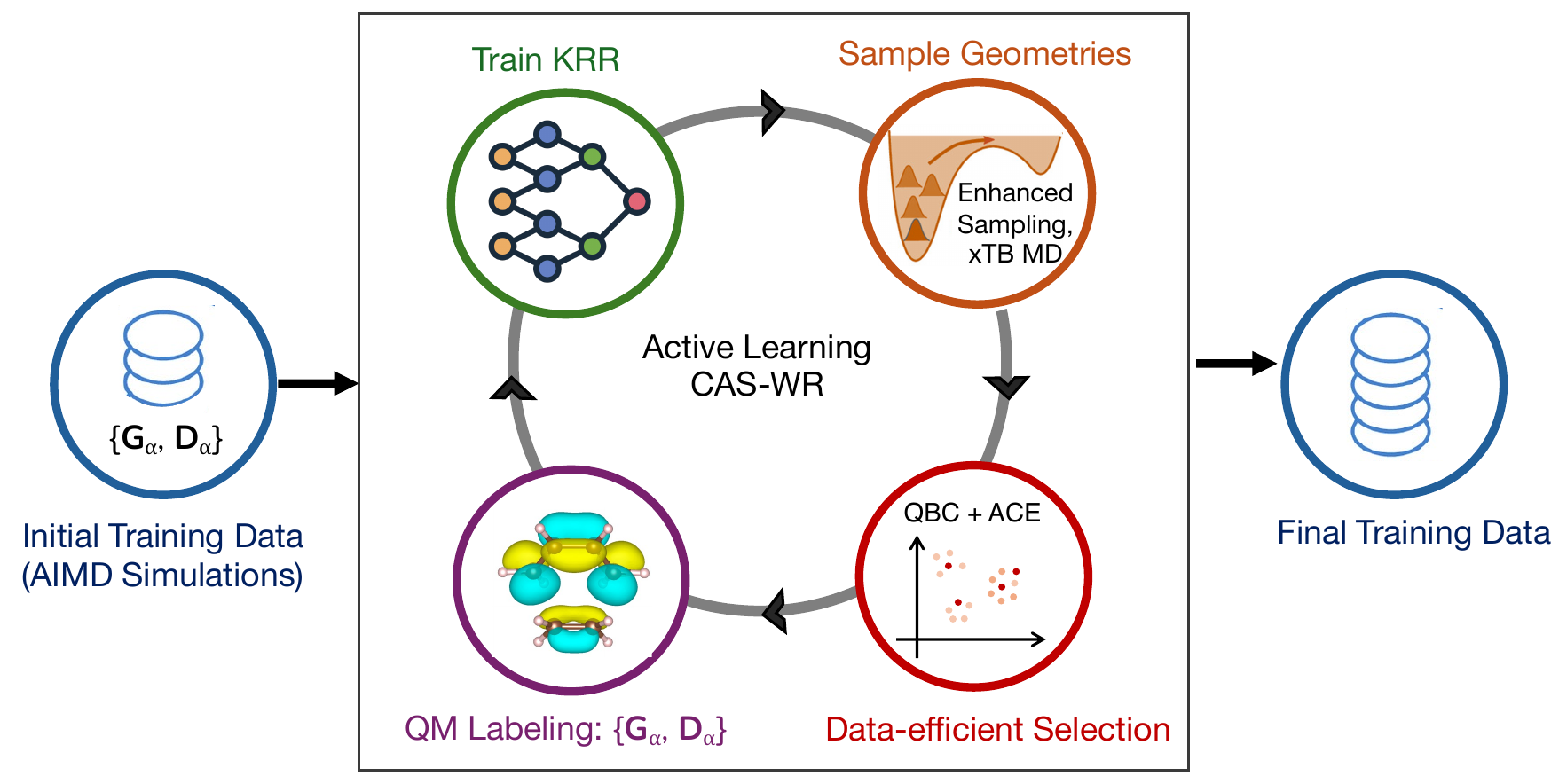}
    \caption{Active learning workflow to train CAS-WR}
    \label{fig:2}
\end{figure}

\begin{figure}[h!]
    \centering
    \includegraphics[width=\linewidth]{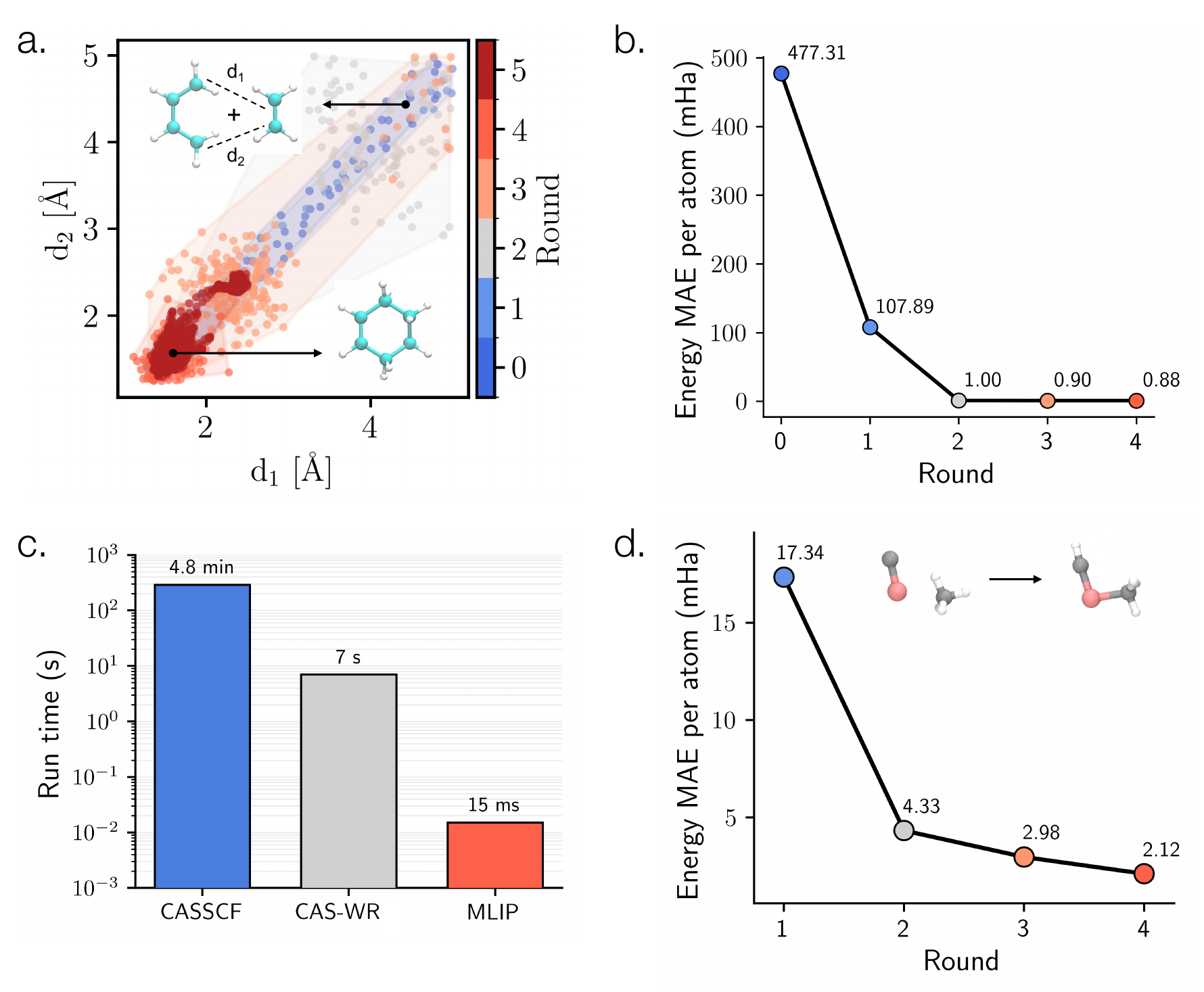}
    \caption{(a) Distribution of configurations selected over successive active-learning rounds for the Diels--Alder cycloaddition of butadiene and ethylene. (b) Mean absolute error of CASCI energies evaluated using the CAS-WR-predicted 1-RDMs as a function of active-learning round for the Diels--Alder system. (\textbf{c}) Comparison of the computational cost of a single energy evaluation using conventional CASSCF, CAS-WR reconstruction, and MACE MLIP inference for the \ce{TiC+}-catalyzed C--H activation of methane. MACE MLIP and CASSCF timings are taken from Seal \textit{et al}.\cite{seal_weighted_2025} (d) CASCI energy errors obtained from CAS-WR-predicted 1-RDMs over successive active-learning rounds for the \ce{TiC+}+\ce{CH4} system.}
    \label{fig:3}
\end{figure}

We next applied CAS-WR to the Diels--Alder cycloaddition of butadiene and ethylene to form cyclohexene, a multidimensional reaction with substantial multiconfigurational character.\cite{foster_ab_2026} The system was modeled using a (6e, 6o) active space comprising its entire $\pi$ system, with MC-PDFT employing the tPBE on-top functional and the def2-TZVP basis set.\cite{seal_computing_2025,mitchell_application_2022} Details of the electronic structure setup are given in Section SII. To construct the training set efficiently, we used a two-stage active-learning protocol. Query-by-committee was first used to identify geometries with large disagreement among independently trained KRR models,\cite{seung_query_1992,schran_committee_2020} thereby selecting configurations with high electronic-structure uncertainty. Gaussian-process diversity screening was then applied in the descriptor space to remove redundant structures.\cite{vandermause_active_2022,perego_data_2024} Further details are provided in Section~SIII of the Supporting Information. The selected configurations were labeled with $\mathbf{D}$ and the model was iteratively retrained. \Cref{fig:2} shows the complete active learning cycle for CAS-WR. The distribution of configurations selected over successive active-learning rounds is shown in \Cref{fig:3}a, while \Cref{fig:3}b reports the corresponding error reduction, quantified by the mean absolute error of the CASCI energies obtained from the CAS-WR predicted 1-RDMs. The error in the predicted 1-RDMs along the active learning rounds are reported in Figure S4.

Next, we demonstrate CAS-WR for an open-shell transition-metal system requiring a multireference description.\cite{geng_intrinsic_2019} The reference data were taken from the active-learning data set previously constructed for the WASP-enabled multireference MLIP describing the \ce{TiC+}-catalyzed C--H activation of methane at the SA(2)-CAS(7e,9o)/def2-TZVP level of theory.\cite{seal_weighted_2025} These calculations already define a consistent active space and CASSCF solution branch across the reactive region, allowing CAS-WR to be trained directly on the same data, as an electronic-structure surrogate model. As shown in \Cref{fig:3}c, CAS-WR is more computationally expensive than a MLIP(MACE\cite{batatia_mace_2022,seal_weighted_2025}) inference, but remains substantially cheaper than repeating CASSCF calculations. Crucially, CAS-WR restores access to arbitrary electronic observables that can be derived from the reconstructed CAS wave function and are not available from a conventional MLIP unless explicitly included as additional learning targets. \Cref{fig:3}d shows the reduction in CAS-WR errors (quantified in terms of CASCI energy errors) over successive active-learning rounds.

\begin{figure}[h!]
    \centering
    \includegraphics[width=\linewidth]{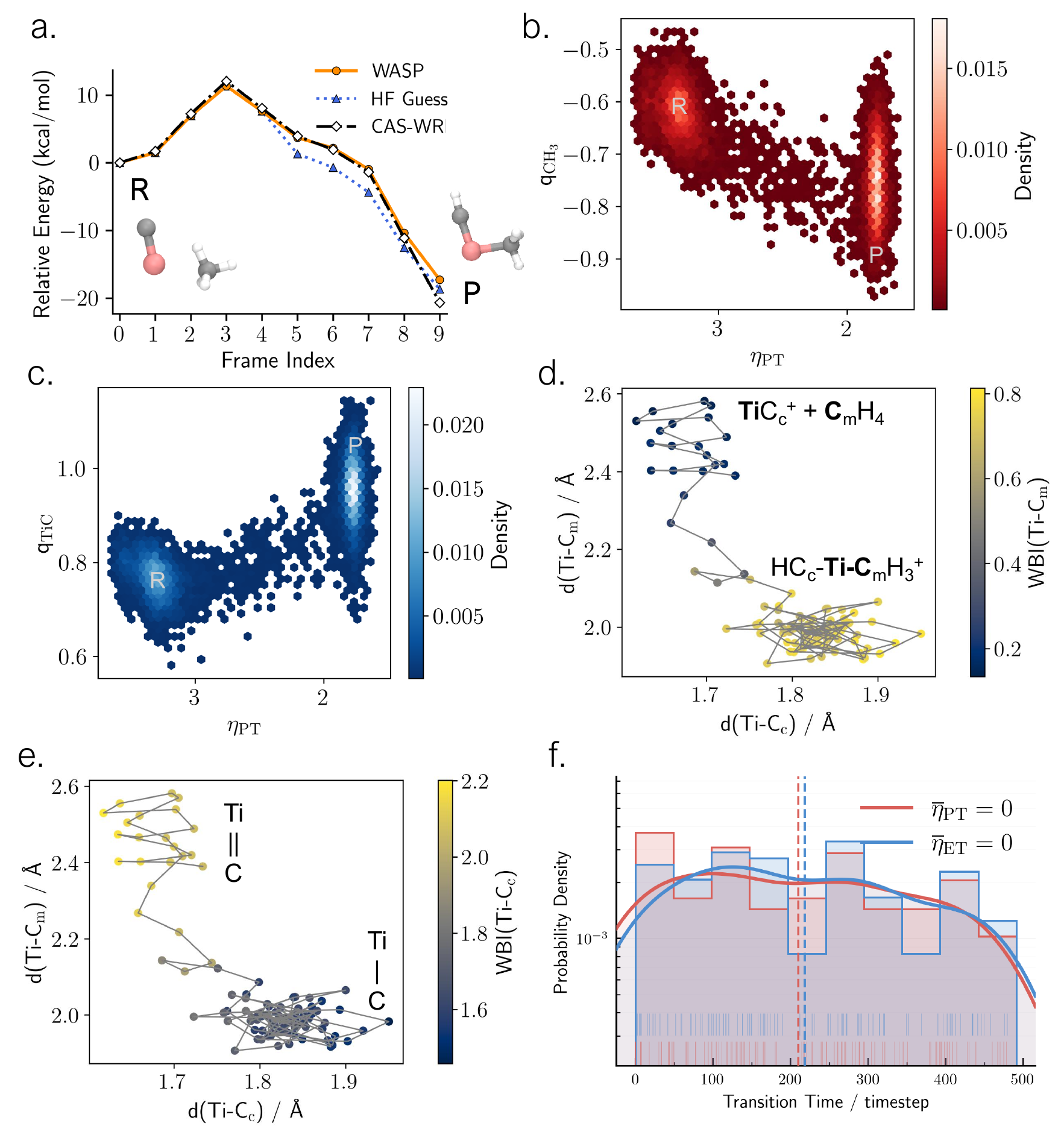}
    \caption{Trajectory-resolved electronic-structure analysis of \ce{TiC+}-catalyzed methane C--H activation using CAS-WR. (a) Comparison of MC-PDFT energies evaluated on top of CASCI wave functions obtained in the CAS-WR-predicted orbital basis with those obtained using independently computed HF orbitals and the corresponding reference energies along the reaction coordinate. (b, c) Fragment-resolved Mulliken charges of the \ce{TiC} and \ce{CH3} moieties. (d, e) Wiberg bond indices describing cleavage of the \ce{Ti-C_{c}} bond and formation of the \ce{Ti-C_{m}} bond over a trajectory. (f) Distribution of the proton-transfer time, $t_{\mathrm{PT}}$ given by $t(\bar{\eta}_{\mathrm{PT}}=0)$ and electron-transfer  times, $t_{\mathrm{ET}}$ given by $t(\bar{\eta}_{\mathrm{ET}}=0)$ over 100 reactive trajectories, where $\bar{\eta}_{\mathrm{PT}}$ and $\bar{\eta}_{\mathrm{ET}}$ denote the corresponding normalized coordinates.}
    \label{fig:4}
\end{figure}

Whereas the previous study focused on the nuclear dynamics of proton-coupled electron transfer (PCET), here we use CAS-WR to resolve the accompanying electron transfer along the reactive trajectories. \Cref{fig:4}a compares the CAS-WR energies with the corresponding MC-PDFT reference values, showing that CASCI in the predicted orbital basis recovers the correct minima along the reaction pathway. We then used CAS-WR to analyze the electronic-structure evolution over 100 reactive trajectories. The transferring hydrogen retains a positive Mulliken charge throughout the reaction, supporting proton transfer, consistent with the mechanism proposed by Geng \textit{et al.}\cite{geng_intrinsic_2019} Fragment-resolved Mulliken charges\cite{mulliken_electronic_1955} further show that the \ce{TiC} moiety becomes more positive while the emerging methyl fragment gains electron density (\Cref{fig:4}b,c), indicating electron transfer from \ce{TiC} to \ce{CH3}.

The bond rearrangement is captured by the Wiberg bond indices.\cite{wiberg_application_1968,mayer_bond_2007} The \ce{Ti-C_{TiC}} bond order decreases, while the \ce{Ti-C_{Me}} bond order increases from approximately zero to one as the metal--methyl bond forms (\Cref{fig:4}d,e). Together with the positive charge on the transferring hydrogen and the increasing electron density on the methyl fragment, these trends support heterolytic cleavage of the methane \ce{C-H} bond, with proton transfer to \ce{C_{TiC}} and electron transfer toward \ce{CH3}. The same mechanistic picture is obtained using intrinsic atomic orbital partitioning evaluated along a representative reactive trajectory (Figure S5).

\clearpage

To quantify the relative timing of proton and electron transfer, we define two collective variables from nuclear ($\eta_{\mathrm{PT}}$) and electronic ($\eta_{\mathrm{ET}}$) degrees of freedom, respectively,
\begin{equation}
\eta_{\mathrm{PT}}=
\mathrm{CN}(\ce{C_{Me}},\ce{H})
-
\mathrm{CN}(\ce{C_{TiC}},\ce{H}),\quad
\eta_{\mathrm{ET}}
=
\text{q}_{\ce{TiC}}
-
\text{q}_{\ce{CH3}},
\end{equation}
where $\mathrm{CN}$ denotes the coordination number and $\text{q}$ is the fragment-resolved charge. The corresponding crossing times provide trajectory-resolved proton- and electron-transfer timescales, allowing synchronous and asynchronous PCET pathways to be distinguished.\cite{cukier_proton-coupled_1998} The distribution in \Cref{fig:4}f indicates that the reaction proceeds through a synchronous PCET mechanism. This illustrates how CAS-WR enables ensemble electronic structure analysis which gives mechanistic insights, inaccessible from the MLIP trajectories alone.

\begin{figure}[h!]
    \centering
    \includegraphics[width=\linewidth]{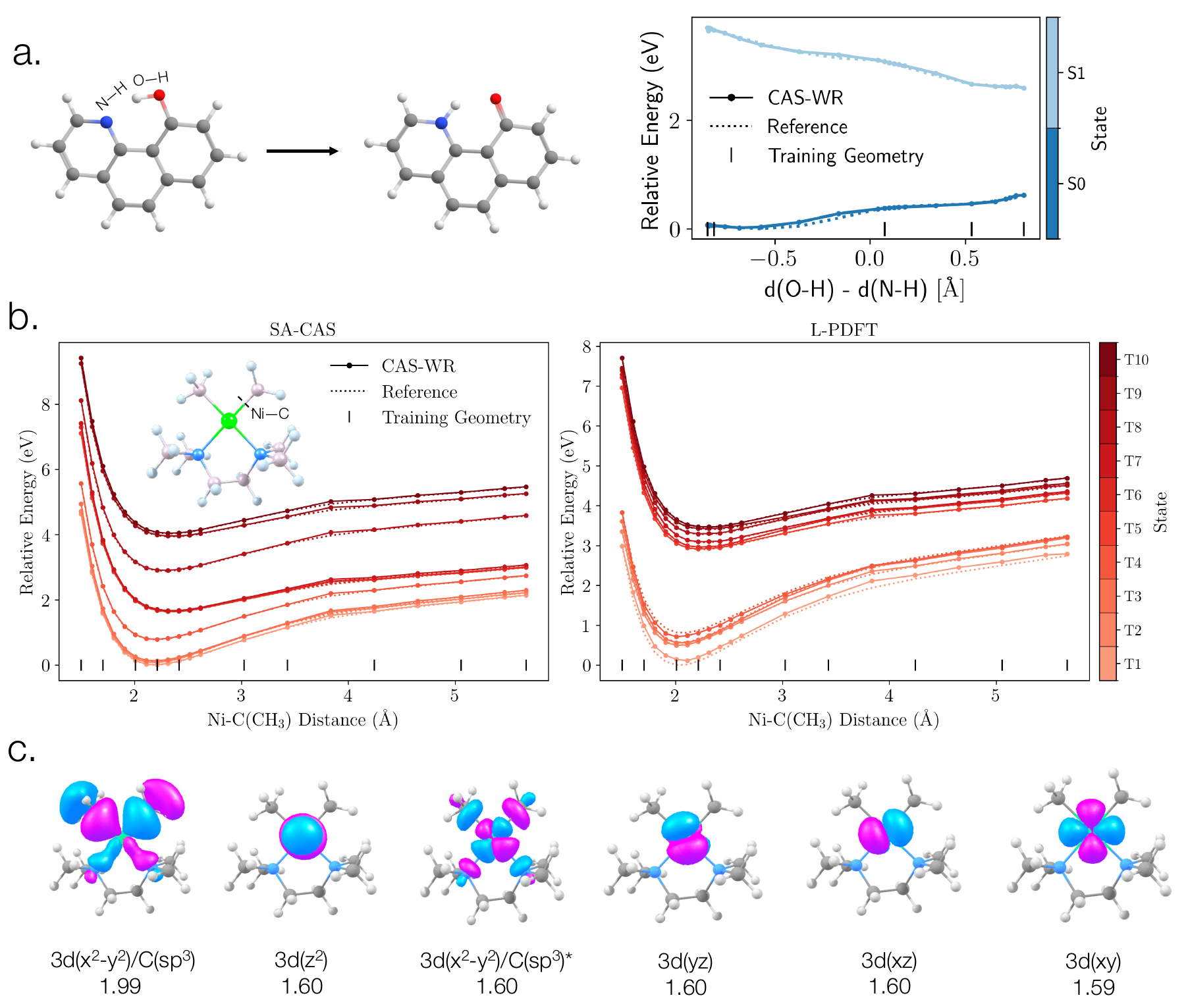}
    \caption{Excited-state reactivity with CAS-WR. (a) L-PDFT potential-energy profiles along the proton-transfer coordinate of 10-hydroxybenzo[h]quinoline. CAS-WR energies are obtained by performing multistate CASCI in the CAS-WR-predicted orbital basis followed by L-PDFT, and are compared with the corresponding reference L-PDFT energies. (b) L-PDFT potential-energy profiles along the \ce{Ni-C(CH3)} bond-dissociation coordinate of the \ce{(TMEDA)Ni^{II}(CH3)2} complex, obtained analogously from multistate CASCI wave functions in the CAS-WR-predicted orbital basis and compared with reference L-PDFT calculations. The inset shows the Ni--C bond-breaking process. The training geometries for both the systems are marked by black dashes. (c) Natural orbitals of the (10e, 6o) active space at the transition state of the \ce{Ni-C} photo-dissociation with the ground-state occupation numbers.}
    \label{fig:5}
\end{figure}

As a preliminary demonstration of CAS-WR for excited-state reactivity, we tested the use of state-averaged 1-RDMs to interpolate excited-state potential-energy cuts. We first considered excited-state intramolecular proton transfer in 10-hydroxybenzo[h]quinoline (HBQ), a system previously studied using linearized-pair density functional theory (L-PDFT),\cite{hennefarth_linearized_2023,clifford_ultrafast_2026} and modeled here with SA(2)-tPBE(4e,4o)/6-31G** along the proton-transfer coordinate (Figure~5a). We then tested CAS-WR for transition-metal photocatalysis, modelling light-activated \ce{C-C} coupling in dialkyl \ce{Ni^{II}} complexes.\cite{cusumano_light-promoted_2025} Using \ce{(TMEDA)Ni^{II}(CH3)2} as a model system, we constructed the potential-energy cut along the \ce{Ni-C(CH3)} stretching coordinate at the SA(10)-tPBE(10e,6o)/6-31G++(d,p) level (Figure~5b). Starting from reference calculations at the endpoints of each reaction coordinate described below, we iteratively added geometries selected to be maximally distinct from those already included in the training set. For both systems, CAS-WR reproduced the full potential-energy profiles using less than 40\% of the total sampled configurations. These results provide an initial demonstration that CAS-WR can be extended beyond ground-state wavefunctions to excited-state reactivity.

In summary, we introduced CAS-WR a machine-learning framework for reconstructing geometry-dependent CAS wave functions without repeated CASSCF orbital optimization. CAS-WR learns the one-electron reduced density matrix of a reference CASSCF solution, uses its natural orbitals to define the active-space orbital basis, and recovers the corresponding multiconfigurational wave function through a subsequent CASCI calculation. This procedure reduces the computational cost by approximately a factor of 30 while preserving access to wavefunction-derived quantities. Among the representations examined, the 1-RDM reproduced the reference CASSCF orbital solution more reliably than the effective Fock matrix for both ground- and excited-state test cases.

For multidimensional reactive systems, we combined CAS-WR with a two-stage active-learning procedure based on 1-RDM uncertainty and structural diversity. Applications to the Diels–Alder reaction and \ce{TiC+}-catalyzed methane activation show that the reconstructed wave functions provide natural-orbital occupations, configuration weights, charge and spin populations, and bond-order measures across complete reactive ensembles. In the methane-activation example, this information identifies the redistribution of charge during C–H cleavage and establishes the synchronous character of the proton-coupled electron-transfer mechanism—information that cannot be obtained from the MLIP trajectory alone. The results therefore demonstrate that CAS-WR goes beyond predicting individual electronic properties: it reconstructs the CAS wave function that can be interrogated through a broad range of electronic observables. Preliminary calculations for excited-state proton transfer and Ni–C photodissociation indicate that CAS-WR can be applied to excited state properties.

Further developments will focus on a differentiable implementation for evaluating gradients,\cite{zhang_differentiable_2022} the calculation of spin–orbit couplings from the reconstructed multistate wave functions,\cite{jangid_linearized_2026} and direct learning of the 1- and 2-RDMs.\cite{b_machine_2026} The latter would remove the CASCI step and permit direct evaluation of one- and two-electron observables and multireference energies. CAS-WR thus provides a route to retaining correlated electronic information across configuration space at a cost compatible with large-scale sampling.

\section{Supporting Information}
Computational details of the electronic-structure calculations and kernel ridge regression models, orbital-overlap quantification, active-learning workflow and associated error statistics, and supplementary figures.

\section{Data Availability}
The CAS-WR code was developed using \texttt{PySCF}\cite{sunPySCF2018,sunRecent2020,sun_python_2026-1} (version 2.12.1) and \texttt{scikit-learn}\cite{pedregosa2011scikit} (version 1.7.2), and is available at \url{https://github.com/aniruddha-seal/cas-wr}.

\section{Acknowledgments}
A.S. acknowledges support from the Benjamin Ball Freud Merit Scholarship Fellowship from the Department of Chemistry at the University of Chicago. A.S., A.L.F., and L.G. acknowledge support from the Catalyst Design for Decarbonization Center, an Energy Frontier Research Center funded by the U.S. Department of Energy, Office of Science, Basic Energy Sciences, under Award No. DE-SC0023383. B.J. and L.G. acknowledge support from the Computational Chemical Sciences Program under Award No. DE-SC0023382, funded by the U.S. Department of Energy, Office of Basic Energy Sciences, the Chemical Sciences, Geosciences, and Biosciences Division. The authors also acknowledge the University of Chicago Research Computing Center for providing computational resources that supported this work.

\section{Conflict of Interest Disclosure}

A.L.F.\ is a co-founder and consultant of Evozyne, Inc.\ and a co-author of US Patent Applications 16/887,710 and 17/642,582, US Provisional Patent Applications 62/853,919, 62/900,420, 63/314,898, 63/479,378, 63/521,617, 63/510,130, 63/669,836, and 63/987,554 and International Patent Applications PCT/US2020/035206, PCT/US2020/050466, PCT/US24/
10805, PCT/US24/34369, and PCT/US25/35833.

\bibliography{ref.bib}

@article{Bannwarth2019GFN2xTB,
    author  = {Bannwarth, Christoph and Ehlert, Sebastian and Grimme, Stefan},
    title   = {{GFN2-xTB}---An Accurate and Broadly Parametrized
               Self-Consistent Tight-Binding Quantum Chemical Method
               with Multipole Electrostatics and Density-Dependent
               Dispersion Contributions},
    journal = {Journal of Chemical Theory and Computation},
    year    = {2019},
    volume  = {15},
    number  = {3},
    pages   = {1652--1671},
    doi     = {10.1021/acs.jctc.8b01176}
  }

@book{murphy2012machine,
  title={Machine learning: a probabilistic perspective},
  author={Murphy, Kevin P},
  year={2012},
  publisher={MIT press}
}

@article{pedregosa2011scikit,
  title={Scikit-learn: Machine learning in Python},
  author={Pedregosa, Fabian and Varoquaux, Ga{\"e}l and Gramfort, Alexandre and Michel, Vincent and Thirion, Bertrand and Grisel, Olivier and Blondel, Mathieu and Prettenhofer, Peter and Weiss, Ron and Dubourg, Vincent and others},
  journal={the Journal of machine Learning research},
  volume={12},
  pages={2825--2830},
  year={2011},
  publisher={JMLR. org}
}

@article{rassolov_6-31g_1998,
	title = {6-{31G}* basis set for atoms {K} through {Zn}},
	volume = {109},
	issn = {0021-9606},
	url = {https://doi.org/10.1063/1.476673},
	doi = {10.1063/1.476673},
	number = {4},
	urldate = {2026-10-12},
	journal = {The Journal of Chemical Physics},
	author = {Rassolov, Vitaly A. and Pople, John A. and Ratner, Mark A. and Windus, Theresa L.},
	month = jul,
	year = {1998},
	pages = {1223--1229},
}

@article{clark_efficient_1983,
	title = {Efficient diffuse function-augmented basis sets for anion calculations. {III}.† {The} 3-21+{G} basis set for first-row elements, {Li}–{F}},
	volume = {4},
	issn = {1096-987X},
	url = {https://onlinelibrary.wiley.com/doi/10.1002/jcc.540040303},
	doi = {10.1002/jcc.540040303},
	language = {en},
	number = {3},
	urldate = {2026-10-12},
	journal = {Journal of Computational Chemistry},
	publisher = {John Wiley \& Sons, Ltd},
	author = {Clark, Timothy and Chandrasekhar, Jayaraman and Spitznagel, Günther W. and Schleyer, Paul Von Ragué},
	month = sep,
	year = {1983},
	pages = {294--301},
}

@article{hehre_selfconsistent_1972,
	title = {Self—{Consistent} {Molecular} {Orbital} {Methods}. {XII}. {Further} {Extensions} of {Gaussian}—{Type} {Basis} {Sets} for {Use} in {Molecular} {Orbital} {Studies} of {Organic} {Molecules}},
	volume = {56},
	issn = {0021-9606},
	url = {https://doi.org/10.1063/1.1677527},
	doi = {10.1063/1.1677527},
	number = {5},
	urldate = {2026-10-12},
	journal = {The Journal of Chemical Physics},
	author = {Hehre, W. J. and Ditchfield, R. and Pople, J. A.},
	month = mar,
	year = {1972},
	pages = {2257--2261},
}

@article{ditchfield_selfconsistent_1971,
	title = {Self‐{Consistent} {Molecular}‐{Orbital} {Methods}. {IX}. {An} {Extended} {Gaussian}‐{Type} {Basis} for {Molecular}‐{Orbital} {Studies} of {Organic} {Molecules}},
	volume = {54},
	issn = {0021-9606},
	url = {https://doi.org/10.1063/1.1674902},
	doi = {10.1063/1.1674902},
	number = {2},
	urldate = {2026-10-12},
	journal = {The Journal of Chemical Physics},
	author = {Ditchfield, R. and Hehre, W. J. and Pople, J. A.},
	month = jan,
	year = {1971},
	pages = {724--728},
}

@article{hariharan_influence_1973,
	title = {The influence of polarization functions on molecular orbital hydrogenation energies},
	volume = {28},
	issn = {1432-2234},
	url = {https://doi.org/10.1007/BF00533485},
	doi = {10.1007/BF00533485},
	language = {en},
	number = {3},
	urldate = {2026-10-12},
	journal = {Theoretica chimica acta},
	author = {Hariharan, P. C. and Pople, J. A.},
	month = sep,
	year = {1973},
	pages = {213--222},
}

@article{sun_python_2026-1,
	title = {The {Python} simulations of chemistry framework: 10 years of an open-source quantum chemistry project},
	volume = {165},
	issn = {0021-9606},
	shorttitle = {The {Python} simulations of chemistry framework},
	url = {https://doi.org/10.1063/5.0337441},
	doi = {10.1063/5.0337441},
	number = {10},
	urldate = {2026-10-12},
	journal = {The Journal of Chemical Physics},
	author = {Sun, Qiming and Hermes, Matthew R. and Wu, Xiaojie and Zhai, Huanchen and Zhang, Xing and Ahmed, Abdelrahman M. and Aucar, Juan José and Backhouse, Oliver J. and Banerjee, Samragni and Bao, Peng and Bogdanov, Nikolay A. and Bystrom, Kyle and Chapoton, Frédéric and Chen, Ning-Yuan and Chernyshov, Ivan Yu. and Clifford, Helen S. and Cohen-Janes, Sander and Cui, Zhi-Hao and Damour, Yann D. and Dattani, Nike and Dittmer, Linus Bjarne and Ehlert, Sebastian and Eriksen, Janus Juul and Evangelista, Francesco A. and Ewing, Simon A. and Farahvash, Ardavan and Focke, Kevin and Gao, Yang and Gasperich, Kevin E. and Gillispie, Nathan and Greiner, Jonas and Hennefarth, Matthew R. and Hermann, Jan and Hillenbrand, Christopher and Huhtasalo, Joonatan and Ibrahim, Basil and Jangid, Bhavnesh and Javaremi, Alireza Nejati and Jenkins, Andrew J. and Jin, Yu and King, Daniel S. and Kooi, Derk Pieter and Kurian, Jo S. and Larsson, Henrik R. and Lau, Bryan Tak Gwong and Lee, Seunghoon and Lehtola, Susi and Li, Chenghan and Li, Hao and Li, Jiachen and Li, Rui and Li, Shuhang and Lykhin, Aleksandr O. and Mahajan, Ankit and Mauger, Nastasia and Mazo-Sevillano, Pablo del and Moussa, Jonathan and Nakano, Kousuke and Neufeld, Verena A. and Peng, Linqing and Pham, Hung Q. and Pinski, Peter and Pokhilko, Pavel and Pu, Zhichen and Qian, Yubing and Quiton, Stephen Jon and Schulze, Wanja T. and Scott, Thais R. and Seal, Aniruddha and Serna, James D. and Smith, James E. T. and Smyser, Kori E. and Stahl, Terrence and Sun, Chong and Sung, Kevin J. and Trushin, Egor and Upadhyay, Shiv and Vo, Ethan A. and Vogels, Thijs and Wang, Shirong and Wang, Tai and Wang, Xiao and Wang, Xubo and Wang, Yuanheng and Williamson, Mark and Yang, Junjie and Ye, Hong-Zhou and Yeh, Chia-Nan and Yu, Haiyang and Yu, Jincheng and Yu, Victor Wen-zhe and Zhang, Chaoqun and Zhang, Dayou and Zhang, Yichi and Zhao, Zijun and Zhou, Zehao and Zhu, Andrew J. and Zhu, Tianyu and Berkelbach, Timothy C. and Gagliardi, Laura and Sharma, Sandeep and Sokolov, Alexander Yu. and Kin-Lic Chan, Garnet},
	month = sep,
	year = {2026},
	pages = {102502},
}

@article{sunPySCF2018,
	author = {Sun, Qiming and Berkelbach, Timothy C. and Blunt, Nick S. and Booth, George H. and Guo, Sheng and Li, Zhendong and Liu, Junzi and McClain, James D. and Sayfutyarova, Elvira R. and Sharma, Sandeep and Wouters, Sebastian and Chan, Garnet Kin-Lic},
	doi = {10.1002/wcms.1340},
	issn = {1759-0884},
	journal = {Wiley Interdiscip. Rev. Comput. Mol. Sci.},
	number = {1},
	pages = {e1340},
	shorttitle = {{{PySCF}}},
	title = {{{PySCF}}: The {{Python-based}} Simulations of Chemistry Framework},
	urldate = {2023-01-12},
	volume = {8},
	year = {2018}}

@article{sunRecent2020,
	author = {Sun, Qiming and Zhang, Xing and Banerjee, Samragni and Bao, Peng and Barbry, Marc and Blunt, Nick S. and Bogdanov, Nikolay A. and Booth, George H. and Chen, Jia and Cui, Zhi-Hao and Eriksen, Janus J. and Gao, Yang and Guo, Sheng and Hermann, Jan and Hermes, Matthew R. and Koh, Kevin and Koval, Peter and Lehtola, Susi and Li, Zhendong and Liu, Junzi and Mardirossian, Narbe and McClain, James D. and Motta, Mario and Mussard, Bastien and Pham, Hung Q. and Pulkin, Artem and Purwanto, Wirawan and Robinson, Paul J. and Ronca, Enrico and Sayfutyarova, Elvira R. and Scheurer, Maximilian and Schurkus, Henry F. and Smith, James E. T. and Sun, Chong and Sun, Shi-Ning and Upadhyay, Shiv and Wagner, Lucas K. and Wang, Xiao and White, Alec and Whitfield, James Daniel and Williamson, Mark J. and Wouters, Sebastian and Yang, Jun and Yu, Jason M. and Zhu, Tianyu and Berkelbach, Timothy C. and Sharma, Sandeep and Sokolov, Alexander Yu. and Chan, Garnet Kin-Lic},
	doi = {10.1063/5.0006074},
	issn = {0021-9606},
	journal = {J. Chem. Phys.},
	month = jul,
	number = {2},
	pages = {024109},
	publisher = {American Institute of Physics},
	title = {Recent Developments in the {{PySCF}} Program Package},
	urldate = {2023-01-12},
	volume = {153},
	year = {2020}}

@article{SunLibcint2015,
	author = {Sun, Qiming},
	copyright = {{\copyright} 2015 Wiley Periodicals, Inc.},
	doi = {10.1002/jcc.23981},
	issn = {1096-987X},
	journal = {J. Comput. Chem.},
	number = {22},
	pages = {1664--1671},
	shorttitle = {Libcint},
	title = {Libcint: {{An}} Efficient General Integral Library for {{Gaussian}} Basis Functions},
	urldate = {2023-10-21},
	volume = {36},
	year = {2015}}

@article{LehtolaRecent2018,
	author = {Lehtola, Susi and Steigemann, Conrad and Oliveira, Micael J. T. and Marques, Miguel A. L.},
	doi = {10.1016/j.softx.2017.11.002},
	issn = {2352-7110},
	journal = {SoftwareX},
	month = jan,
	pages = {1--5},
	publisher = {Elsevier},
	title = {Recent Developments in Libxc --- {{A}} Comprehensive Library of Functionals for Density Functional Theory},
	urldate = {2023-10-21},
	volume = {7},
	year = {2018}}

@article{MarquesLibxc2012,
	author = {Marques, Miguel A. L. and Oliveira, Micael J. T. and Burnus, Tobias},
	doi = {10.1016/j.cpc.2012.05.007},
	issn = {0010-4655},
	journal = {Comput. Phys. Commun.},
	month = oct,
	number = {10},
	pages = {2272--2281},
	shorttitle = {Libxc},
	title = {Libxc: {{A}} Library of Exchange and Correlation Functionals for Density Functional Theory},
	urldate = {2023-10-21},
	volume = {183},
	year = {2012}}

@misc{b_machine_2026,
	title = {Machine learning the two-electron reduced density matrix in molecules and condensed phases},
	url = {http://arxiv.org/abs/2603.06882},
	doi = {10.48550/arXiv.2603.06882},
	urldate = {2026-09-03},
	publisher = {arXiv},
	author = {B, Jessica A. Martinez and Rana, Bhaskar and Shao, Xuecheng and Pernal, Katarzyna and Pavanello, Michele},
	month = mar,
	year = {2026},
	note = {arXiv:2603.06882 [physics.chem-ph]},
}

@article{yang_reactant-induced_2023,
	title = {Reactant-induced dynamics of lithium imide surfaces during the ammonia decomposition process},
	volume = {6},
	copyright = {2023 The Author(s), under exclusive licence to Springer Nature Limited},
	issn = {2520-1158},
	url = {https://www.nature.com/articles/s41929-023-01006-2},
	doi = {10.1038/s41929-023-01006-2},
	language = {en},
	number = {9},
	urldate = {2026-09-03},
	journal = {Nature Catalysis},
	publisher = {Nature Publishing Group},
	author = {Yang, Manyi and Raucci, Umberto and Parrinello, Michele},
	month = sep,
	year = {2023},
	pages = {829--836},
}

@article{zhang_differentiable_2022,
	title = {Differentiable quantum chemistry with {PySCF} for molecules and materials at the mean-field level and beyond},
	volume = {157},
	issn = {0021-9606},
	url = {https://doi.org/10.1063/5.0118200},
	doi = {10.1063/5.0118200},
	number = {20},
	urldate = {2026-09-03},
	journal = {The Journal of Chemical Physics},
	author = {Zhang, Xing and Chan, Garnet Kin-Lic},
	month = nov,
	year = {2022},
	pages = {204801},
}

@article{hennefarth_linearized_2023,
	title = {Linearized {Pair}-{Density} {Functional} {Theory}},
	volume = {19},
	issn = {1549-9618},
	url = {https://doi.org/10.1021/acs.jctc.3c00207},
	doi = {10.1021/acs.jctc.3c00207},
	number = {11},
	urldate = {2026-09-03},
	journal = {Journal of Chemical Theory and Computation},
	author = {Hennefarth, Matthew R. and Hermes, Matthew R. and Truhlar, Donald G. and Gagliardi, Laura},
	month = may,
	year = {2023},
	pages = {3172--3183},
}

@article{cusumano_light-promoted_2025,
	title = {Light-{Promoted} {C}(sp3)–{C}(sp3) {Reductive} {Elimination} from {Dialkyl} {NiII} {Complexes}},
	volume = {147},
	issn = {0002-7863},
	url = {https://doi.org/10.1021/jacs.5c09925},
	doi = {10.1021/jacs.5c09925},
	number = {36},
	urldate = {2026-09-03},
	journal = {Journal of the American Chemical Society},
	author = {Cusumano, Alexander Q. and Chaffin, Braden C. and Cagan, David A. and DiLuzio, Stephen and Sutcliffe, Erica and Hadt, Ryan G. and Doyle, Abigail G.},
	month = aug,
	year = {2025},
	pages = {32941--32950},
}

@article{clifford_ultrafast_2026,
	title = {Ultrafast excited-state proton transfer dynamics using linearized pair-density functional theory},
	volume = {17},
	issn = {2041-6520},
	url = {https://doi.org/10.1039/d6sc01160h},
	doi = {10.1039/d6sc01160h},
	number = {32},
	urldate = {2026-09-03},
	journal = {Chemical Science},
	author = {Clifford, Helen S. and Seal, Aniruddha and Gagliardi, Laura},
	month = aug,
	year = {2026},
	pages = {15607--15615},
}

@article{mulliken_electronic_1955,
	title = {Electronic {Population} {Analysis} on {LCAO}–{MO} {Molecular} {Wave} {Functions}. {I}},
	volume = {23},
	issn = {0021-9606},
	url = {https://doi.org/10.1063/1.1740588},
	doi = {10.1063/1.1740588},
	number = {10},
	urldate = {2026-09-03},
	journal = {The Journal of Chemical Physics},
	author = {Mulliken, R. S.},
	month = oct,
	year = {1955},
	pages = {1833--1840},
}

@article{mayer_bond_2007,
	title = {Bond order and valence indices: {A} personal account},
	volume = {28},
	copyright = {Copyright © 2006 Wiley Periodicals, Inc.},
	issn = {1096-987X},
	shorttitle = {Bond order and valence indices},
	url = {https://onlinelibrary.wiley.com/doi/abs/10.1002/jcc.20494},
	doi = {10.1002/jcc.20494},
	language = {de},
	number = {1},
	urldate = {2026-09-03},
	journal = {Journal of Computational Chemistry},
	author = {Mayer, I.},
	year = {2007},
	pages = {204--221},
}

@article{wiberg_application_1968,
	title = {Application of the pople-santry-segal {CNDO} method to the cyclopropylcarbinyl and cyclobutyl cation and to bicyclobutane},
	volume = {24},
	issn = {0040-4020},
	url = {https://www.sciencedirect.com/science/article/pii/0040402068880573},
	doi = {10.1016/0040-4020(68)88057-3},
	number = {3},
	urldate = {2026-09-03},
	journal = {Tetrahedron},
	author = {Wiberg, K. B.},
	month = jan,
	year = {1968},
	pages = {1083--1096},
}

@article{cukier_proton-coupled_1998,
	title = {Proton-{Coupled} {Electron} {Transfer}},
	volume = {49},
	issn = {1545-1593},
	url = {https://www.annualreviews.org/content/journals/10.1146/annurev.physchem.49.1.337},
	doi = {10.1146/annurev.physchem.49.1.337},
	urldate = {2026-09-03},
	journal = {Annual Review of Physical Chemistry},
	publisher = {Annual Reviews},
	author = {Cukier, Robert I. and Nocera, Daniel G.},
	year = {1998},
	pages = {337--369},
}

@book{hastie_elements_2009,
	address = {New York, NY},
	series = {Springer {Series} in {Statistics}},
	title = {The {Elements} of {Statistical} {Learning}},
	copyright = {http://www.springer.com/tdm},
	isbn = {978-0-387-84857-0 978-0-387-84858-7},
	url = {http://link.springer.com/10.1007/978-0-387-84858-7},
	doi = {10.1007/978-0-387-84858-7},
	urldate = {2026-09-03},
	publisher = {Springer},
	author = {Hastie, Trevor and Tibshirani, Robert and Friedman, Jerome},
	year = {2009},
}

@article{vandermause_active_2022,
	title = {Active learning of reactive {Bayesian} force fields applied to heterogeneous catalysis dynamics of {H}/{Pt}},
	volume = {13},
	copyright = {2022 The Author(s)},
	issn = {2041-1723},
	url = {https://www.nature.com/articles/s41467-022-32294-0},
	doi = {10.1038/s41467-022-32294-0},
	language = {en},
	number = {1},
	urldate = {2026-09-03},
	journal = {Nature Communications},
	publisher = {Nature Publishing Group},
	author = {Vandermause, Jonathan and Xie, Yu and Lim, Jin Soo and Owen, Cameron J. and Kozinsky, Boris},
	month = sep,
	year = {2022},
	pages = {5183},
}

@inproceedings{seung_query_1992,
	address = {New York, NY, USA},
	series = {{COLT} '92},
	title = {Query by committee},
	isbn = {978-0-89791-497-0},
	url = {https://dl.acm.org/doi/10.1145/130385.130417},
	doi = {10.1145/130385.130417},
	urldate = {2026-09-03},
	booktitle = {Proceedings of the fifth annual workshop on {Computational} learning theory},
	publisher = {Association for Computing Machinery},
	author = {Seung, H. S. and Opper, M. and Sompolinsky, H.},
	month = jul,
	year = {1992},
	pages = {287--294},
}

@article{schran_committee_2020,
	title = {Committee neural network potentials control generalization errors and enable active learning},
	volume = {153},
	issn = {0021-9606},
	url = {https://doi.org/10.1063/5.0016004},
	doi = {10.1063/5.0016004},
	number = {10},
	urldate = {2026-09-03},
	journal = {The Journal of Chemical Physics},
	author = {Schran, Christoph and Brezina, Krystof and Marsalek, Ondrej},
	month = sep,
	year = {2020},
	pages = {104105},
}

@article{rana_learning_2025,
	title = {Learning the {One}-{Electron} {Reduced} {Density} {Matrix} at {SCF} {Convergence} {Thresholds}},
	volume = {21},
	issn = {1549-9618},
	url = {https://doi.org/10.1021/acs.jctc.5c01564},
	doi = {10.1021/acs.jctc.5c01564},
	number = {24},
	urldate = {2026-09-03},
	journal = {Journal of Chemical Theory and Computation},
	author = {Rana, Bhaskar and Viot, Nicolas and Martinez B, Jessica A. and Shao, Xuecheng and Ramos, Pablo and Pavanello, Michele},
	month = dec,
	year = {2025},
	pages = {12652--12663},
}

@article{nigam_machine_2026,
	title = {Machine learning of electronic structure and atomistic properties from the external potential},
	volume = {165},
	issn = {0021-9606},
	url = {https://doi.org/10.1063/5.0332678},
	doi = {10.1063/5.0332678},
	number = {3},
	urldate = {2026-09-03},
	journal = {The Journal of Chemical Physics},
	author = {Nigam, Jigyasa and Smidt, Tess and Dusson, Geneviève},
	month = jul,
	year = {2026},
	pages = {034119},
}

@article{zhang_multiconfigurational_2020,
	title = {Multiconfigurational {Effects} on the {Density} {Coherence}},
	volume = {16},
	issn = {1549-9618},
	url = {https://doi.org/10.1021/acs.jctc.0c00903},
	doi = {10.1021/acs.jctc.0c00903},
	number = {11},
	urldate = {2026-09-03},
	journal = {Journal of Chemical Theory and Computation},
	author = {Zhang, Dayou and Truhlar, Donald G.},
	month = oct,
	year = {2020},
	pages = {6915--6925},
}

@misc{van_workum_casnet_2023,
	title = {{CASNet}: {Learning} {Complete} {Active} {Space} {Orbitals} using {Message} {Passing} {Neural} {Networks}},
	copyright = {https://creativecommons.org/licenses/by-nc-nd/4.0/},
	shorttitle = {{CASNet}},
	url = {https://chemrxiv.org/doi/full/10.26434/chemrxiv-2023-lwj87},
	doi = {10.26434/chemrxiv-2023-lwj87},
	language = {en},
	urldate = {2026-09-03},
	author = {Van Workum, Ruard and Malhado, Joao and Marquetand, Philipp},
	month = jun,
	year = {2023},
}

@article{bearpark_can_1996,
	title = {Can {Fulvene} {S1} {Decay} {Be} {Controlled}? {A} {CASSCF} {Study} with {MMVB} {Dynamics}},
	volume = {118},
	issn = {0002-7863},
	shorttitle = {Can {Fulvene} {S1} {Decay} {Be} {Controlled}?},
	url = {https://doi.org/10.1021/ja9542799},
	doi = {10.1021/ja9542799},
	number = {22},
	urldate = {2026-09-03},
	journal = {Journal of the American Chemical Society},
	author = {Bearpark, Michael J. and Bernardi, Fernando and Olivucci, Massimo and Robb, Michael A. and Smith, Barry R.},
	month = jun,
	year = {1996},
	pages = {5254--5260},
}

@book{cramer_essentials_2013,
	title = {Essentials of {Computational} {Chemistry}: {Theories} and {Models}},
	isbn = {978-1-118-71227-6},
	shorttitle = {Essentials of {Computational} {Chemistry}},
	language = {en},
	publisher = {John Wiley \& Sons},
	author = {Cramer, Christopher J.},
	month = apr,
	year = {2013},
	note = {Google-Books-ID: k4R6cf7I7q0C},
}

@article{wardzala_multireference_2026,
	title = {Multireference {Methods} for {Chemistry} and {Materials} {Science}: {Automated} {Active} {Spaces}, {Efficient} {Dynamic} {Correlation}, and {Extended} {Systems}},
	volume = {126},
	issn = {0009-2665},
	shorttitle = {Multireference {Methods} for {Chemistry} and {Materials} {Science}},
	url = {https://doi.org/10.1021/acs.chemrev.5c00866},
	doi = {10.1021/acs.chemrev.5c00866},
	number = {8},
	urldate = {2026-09-03},
	journal = {Chemical Reviews},
	author = {Wardzala, Jacob J. and Hennefarth, Matthew R. and Agarawal, Valay and Jangid, Bhavnesh and Seal, Aniruddha and Hermes, Matthew R. and King, Daniel S. and Gagliardi, Laura},
	month = apr,
	year = {2026},
	pages = {4592--4618},
}

@article{foster_ab_2026,
	title = {An ab initio foundation model of wavefunctions that accurately describes chemical bond breaking},
	copyright = {2026 The Author(s)},
	issn = {2041-1723},
	url = {https://www.nature.com/articles/s41467-026-76604-2},
	doi = {10.1038/s41467-026-76604-2},
	language = {en},
	urldate = {2026-09-03},
	journal = {Nature Communications},
	publisher = {Nature Publishing Group},
	author = {Foster, Adam and Schätzle, Zeno and Szabó, P. Bernát and Cheng, Lixue and Köhler, Jonas and Cassella, Gino and Gao, Nicholas and Li, Jiawei and Noé, Frank and Hermann, Jan},
	month = aug,
	year = {2026},
}

@article{cignoni_electronic_2024,
	title = {Electronic {Excited} {States} from {Physically} {Constrained} {Machine} {Learning}},
	volume = {10},
	issn = {2374-7943},
	url = {https://doi.org/10.1021/acscentsci.3c01480},
	doi = {10.1021/acscentsci.3c01480},
	number = {3},
	urldate = {2026-09-03},
	journal = {ACS Central Science},
	author = {Cignoni, Edoardo and Suman, Divya and Nigam, Jigyasa and Cupellini, Lorenzo and Mennucci, Benedetta and Ceriotti, Michele},
	month = feb,
	year = {2024},
	pages = {637--648},
}

@article{grisafi_transferable_2018,
	title = {Transferable {Machine}-{Learning} {Model} of the {Electron} {Density}},
	volume = {5},
	issn = {2374-7943},
	url = {https://doi.org/10.1021/acscentsci.8b00551},
	doi = {10.1021/acscentsci.8b00551},
	number = {1},
	urldate = {2026-09-03},
	journal = {ACS Central Science},
	author = {Grisafi, Andrea and Fabrizio, Alberto and Meyer, Benjamin and Wilkins, David M. and Corminboeuf, Clemence and Ceriotti, Michele},
	month = dec,
	year = {2018},
	pages = {57--64},
}

@article{shao_machine_2023,
	title = {Machine learning electronic structure methods based on the one-electron reduced density matrix},
	volume = {14},
	copyright = {2023 The Author(s)},
	issn = {2041-1723},
	url = {https://www.nature.com/articles/s41467-023-41953-9},
	doi = {10.1038/s41467-023-41953-9},
	language = {en},
	number = {1},
	urldate = {2026-09-03},
	journal = {Nature Communications},
	publisher = {Nature Publishing Group},
	author = {Shao, Xuecheng and Paetow, Lukas and Tuckerman, Mark E. and Pavanello, Michele},
	month = oct,
	year = {2023},
	pages = {6281},
}

@article{pfau_ab_2020,
	title = {\textit{{Ab} initio} solution of the many-electron {Schrödinger} equation with deep neural networks},
	volume = {2},
	doi = {10.1103/PhysRevResearch.2.033429},
	number = {3},
	journal = {Physical Review Research},
	author = {Pfau, David},
	year = {2020},
}

@article{hermann_deep-neural-network_2020,
	title = {Deep-neural-network solution of the electronic {Schrödinger} equation},
	volume = {12},
	copyright = {2020 The Author(s), under exclusive licence to Springer Nature Limited},
	issn = {1755-4349},
	url = {https://www.nature.com/articles/s41557-020-0544-y},
	doi = {10.1038/s41557-020-0544-y},
	language = {en},
	number = {10},
	urldate = {2026-09-03},
	journal = {Nature Chemistry},
	publisher = {Nature Publishing Group},
	author = {Hermann, Jan and Schätzle, Zeno and Noé, Frank},
	month = oct,
	year = {2020},
	pages = {891--897},
}

@article{pfau_accurate_2024,
	title = {Accurate computation of quantum excited states with neural networks},
	volume = {385},
	url = {https://www.science.org/doi/abs/10.1126/science.adn0137},
	doi = {10.1126/science.adn0137},
	number = {6711},
	urldate = {2026-09-03},
	journal = {Science},
	publisher = {American Association for the Advancement of Science},
	author = {Pfau, David and Axelrod, Simon and Sutterud, Halvard and von Glehn, Ingrid and Spencer, James S.},
	month = aug,
	year = {2024},
	pages = {eadn0137},
}

@article{gong_general_2023,
	title = {General framework for {E}(3)-equivariant neural network representation of density functional theory {Hamiltonian}},
	volume = {14},
	copyright = {2023 The Author(s)},
	issn = {2041-1723},
	url = {https://www.nature.com/articles/s41467-023-38468-8},
	doi = {10.1038/s41467-023-38468-8},
	language = {en},
	number = {1},
	urldate = {2026-09-03},
	journal = {Nature Communications},
	publisher = {Nature Publishing Group},
	author = {Gong, Xiaoxun and Li, He and Zou, Nianlong and Xu, Runzhang and Duan, Wenhui and Xu, Yong},
	month = may,
	year = {2023},
	pages = {2848},
}

@article{schutt_unifying_2019,
	title = {Unifying machine learning and quantum chemistry with a deep neural network for molecular wavefunctions},
	volume = {10},
	copyright = {2019 The Author(s)},
	issn = {2041-1723},
	url = {https://www.nature.com/articles/s41467-019-12875-2},
	doi = {10.1038/s41467-019-12875-2},
	language = {en},
	number = {1},
	urldate = {2026-09-03},
	journal = {Nature Communications},
	publisher = {Nature Publishing Group},
	author = {Schütt, K. T. and Gastegger, M. and Tkatchenko, A. and Müller, K.-R. and Maurer, R. J.},
	month = nov,
	year = {2019},
	pages = {5024},
}

@article{hindges_recent_2026,
	title = {Recent advances in electronic structure learning},
	volume = {34},
	issn = {0965-0393},
	url = {https://doi.org/10.1088/1361-651X/ae72ac},
	doi = {10.1088/1361-651X/ae72ac},
	language = {en},
	number = {5},
	urldate = {2026-09-03},
	journal = {Modelling and Simulation in Materials Science and Engineering},
	publisher = {IOP Publishing},
	author = {Hindges, Tim S and He, Wenhao and Li, Ju},
	month = jul,
	year = {2026},
	pages = {053001},
}

@article{tang_deep-learning_2025,
	title = {Deep-learning electronic structure calculations},
	volume = {5},
	copyright = {2025 Springer Nature America, Inc.},
	issn = {2662-8457},
	url = {https://www.nature.com/articles/s43588-025-00932-4},
	doi = {10.1038/s43588-025-00932-4},
	language = {en},
	number = {12},
	urldate = {2026-09-03},
	journal = {Nature Computational Science},
	publisher = {Nature Publishing Group},
	author = {Tang, Zechen and Chen, Haoxiang and Li, Yang and Qian, Yubing and Wang, Yuxiang and Fu, Weizhong and Li, Jialin and Si, Chen and Duan, Wenhui and Chen, Ji and Xu, Yong},
	month = dec,
	year = {2025},
	pages = {1133--1146},
}

@article{yan_atomistic_2026,
	title = {Atomistic simulation study of diamond doping based on machine learning potential},
	volume = {28},
	issn = {1463-9076},
	url = {https://doi.org/10.1039/d5cp03568f},
	doi = {10.1039/d5cp03568f},
	number = {1},
	urldate = {2026-09-03},
	journal = {Physical Chemistry Chemical Physics},
	author = {Yan, Yiheng and Yu, Yaochen and Hu, Junwei and Sun, Xuecheng and Luo, Qinlan and Chu, Zengyong and Dai, Jiayu and Niu, Haiyang},
	month = jan,
	year = {2026},
	pages = {397--407},
}

@article{cui_photocatalyst_2025,
	title = {Photocatalyst {Design} with an {Electrical}-{Polarization}-{Ability} {Descriptor} for {Exclusive} {C}–{H} {Bond} {Activation} in {Methanol}},
	volume = {147},
	issn = {0002-7863},
	url = {https://doi.org/10.1021/jacs.5c02280},
	doi = {10.1021/jacs.5c02280},
	number = {26},
	urldate = {2026-09-03},
	journal = {Journal of the American Chemical Society},
	author = {Cui, Leyuan and Feng, Jun and Wu, Shimiao and Xu, Sha and Wu, Xuejiao and Zhang, Shuhong and Zhang, Qinghong and Fu, Gang and Wang, Ye and Xie, Shunji},
	month = jun,
	year = {2025},
	pages = {22509--22516},
}

@article{wang_eg_2019,
	title = {eg occupancy as an effective descriptor for the catalytic activity of perovskite oxide-based peroxidase mimics},
	volume = {10},
	copyright = {2019 The Author(s)},
	issn = {2041-1723},
	url = {https://www.nature.com/articles/s41467-019-08657-5},
	doi = {10.1038/s41467-019-08657-5},
	language = {en},
	number = {1},
	urldate = {2026-09-03},
	journal = {Nature Communications},
	publisher = {Nature Publishing Group},
	author = {Wang, Xiaoyu and Gao, Xuejiao J. and Qin, Li and Wang, Changda and Song, Li and Zhou, Yong-Ning and Zhu, Guoyin and Cao, Wen and Lin, Shichao and Zhou, Liqi and Wang, Kang and Zhang, Huigang and Jin, Zhong and Wang, Peng and Gao, Xingfa and Wei, Hui},
	month = feb,
	year = {2019},
	pages = {704},
}

@article{zhang_spin_2024,
	title = {Spin states of metal centers in electrocatalysis},
	volume = {53},
	issn = {0306-0012},
	url = {https://doi.org/10.1039/d3cs00913k},
	doi = {10.1039/d3cs00913k},
	number = {16},
	urldate = {2026-09-03},
	journal = {Chemical Society Reviews},
	author = {Zhang, Yuwei and Wu, Qian and Seow, Justin Zhu Yeow and Jia, Yingjie and Ren, Xiao and Xu, Zhichuan J.},
	month = aug,
	year = {2024},
	pages = {8123--8136},
}

@article{suntivich_design_2011,
	title = {Design principles for oxygen-reduction activity on perovskite oxide catalysts for fuel cells and metal–air batteries},
	volume = {3},
	copyright = {2011 Springer Nature Limited},
	issn = {1755-4349},
	url = {https://www.nature.com/articles/nchem.1069},
	doi = {10.1038/nchem.1069},
	language = {en},
	number = {7},
	urldate = {2026-09-03},
	journal = {Nature Chemistry},
	author = {Suntivich, Jin and Gasteiger, Hubert A. and Yabuuchi, Naoaki and Nakanishi, Haruyuki and Goodenough, John B. and Shao-Horn, Yang},
	month = jul,
	year = {2011},
	pages = {546--550},
}

@article{deringer_machine_2019,
	title = {Machine {Learning} {Interatomic} {Potentials} as {Emerging} {Tools} for {Materials} {Science}},
	volume = {31},
	issn = {1521-4095},
	url = {https://onlinelibrary.wiley.com/doi/abs/10.1002/adma.201902765},
	doi = {10.1002/adma.201902765},
	language = {en},
	number = {46},
	urldate = {2026-08-20},
	journal = {Advanced Materials},
	author = {Deringer, Volker L. and Caro, Miguel A. and Csányi, Gábor},
	year = {2019},
	pages = {1902765},
}

@article{unke_machine_2021,
	title = {Machine {Learning} {Force} {Fields}},
	volume = {121},
	issn = {0009-2665},
	url = {https://doi.org/10.1021/acs.chemrev.0c01111},
	doi = {10.1021/acs.chemrev.0c01111},
	number = {16},
	urldate = {2026-08-20},
	journal = {Chemical Reviews},
	author = {Unke, Oliver T. and Chmiela, Stefan and Sauceda, Huziel E. and Gastegger, Michael and Poltavsky, Igor and Schütt, Kristof T. and Tkatchenko, Alexandre and Müller, Klaus-Robert},
	month = mar,
	year = {2021},
	pages = {10142--10186},
}

@article{seal_computing_2025,
	title = {Computing {Reaction} {Kinetics} with {MC}-{PDFT}–{OPESf}: {Combining} {Multireference} {Electronic} {Structure} {Theory} and {Enhanced} {Sampling}},
	volume = {16},
	shorttitle = {Computing {Reaction} {Kinetics} with {MC}-{PDFT}–{OPESf}},
	url = {https://doi.org/10.1021/acs.jpclett.5c02966},
	doi = {10.1021/acs.jpclett.5c02966},
	number = {44},
	urldate = {2025-12-08},
	journal = {The Journal of Physical Chemistry Letters},
	publisher = {American Chemical Society},
	author = {Seal, Aniruddha and Gagliardi, Laura and Ferguson, Andrew L.},
	month = nov,
	year = {2025},
	pages = {11458--11463},
}

@article{mitchell_application_2022,
	title = {Application of {Multiconfiguration} {Pair}-{Density} {Functional} {Theory} to the {Diels}–{Alder} {Reaction}},
	volume = {126},
	issn = {1089-5639},
	url = {https://doi.org/10.1021/acs.jpca.2c06433},
	doi = {10.1021/acs.jpca.2c06433},
	number = {47},
	urldate = {2025-05-13},
	journal = {The Journal of Physical Chemistry A},
	author = {Mitchell, Erica C. and Scott, Thais R. and Bao, Jie J. and Truhlar, Donald G.},
	month = dec,
	year = {2022},
	pages = {8834--8843},
}

@article{seal_weighted_2025,
	title = {Weighted active space protocol for multireference machine-learned potentials},
	volume = {122},
	url = {https://doi.org/10.1073/pnas.2513693122},
	doi = {10.1073/pnas.2513693122},
	number = {38},
	urldate = {2025-10-02},
	journal = {Proceedings of the National Academy of Sciences},
	author = {Seal, Aniruddha and Perego, Simone and Hennefarth, Matthew R. and Raucci, Umberto and Bonati, Luigi and Ferguson, Andrew L. and Parrinello, Michele and Gagliardi, Laura},
	month = sep,
	year = {2025},
	pages = {e2513693122},
}

@article{rath_interpolating_2025,
	title = {Interpolating numerically exact many-body wave functions for accelerated molecular dynamics},
	volume = {16},
	issn = {2041-1723},
	url = {https://doi.org/10.1038/s41467-025-57134-9},
	doi = {10.1038/s41467-025-57134-9},
	number = {1},
	journal = {Nature Communications},
	author = {Rath, Yannic and Booth, George H.},
	month = feb,
	year = {2025},
	pages = {2005},
}

@article{batatia_mace_2022,
	title = {{MACE}: {Higher} {Order} {Equivariant} {Message} {Passing} {Neural} {Networks} for {Fast} and {Accurate} {Force} {Fields}},
	volume = {35},
	shorttitle = {{MACE}},
	url = {https://proceedings.neurips.cc/paper_files/paper/2022/hash/4a36c3c51af11ed9f34615b81edb5bbc-Abstract-Conference.html},
	urldate = {2025-01-08},
	journal = {Advances in Neural Information Processing Systems},
	author = {Batatia, Ilyes and Kovacs, David P. and Simm, Gregor and Ortner, Christoph and Csanyi, Gabor},
	month = dec,
	year = {2022},
	pages = {11423--11436},
}

@article{schran_machine_2021,
	title = {Machine learning potentials for complex aqueous systems made simple},
	volume = {118},
	url = {https://www.pnas.org/doi/10.1073/pnas.2110077118},
	doi = {10.1073/pnas.2110077118},
	number = {38},
	urldate = {2025-05-15},
	journal = {Proceedings of the National Academy of Sciences},
	author = {Schran, Christoph and Thiemann, Fabian L. and Rowe, Patrick and Müller, Erich A. and Marsalek, Ondrej and Michaelides, Angelos},
	month = sep,
	year = {2021},
	pages = {e2110077118},
}

@article{weigend_balanced_2005,
	title = {Balanced basis sets of split valence, triple zeta valence and quadruple zeta valence quality for {H} to {Rn}: {Design} and assessment of accuracy},
	volume = {7},
	issn = {1463-9084},
	shorttitle = {Balanced basis sets of split valence, triple zeta valence and quadruple zeta valence quality for {H} to {Rn}},
	url = {https://pubs.rsc.org/en/content/articlelanding/2005/cp/b508541a},
	doi = {10.1039/B508541A},
	language = {en},
	number = {18},
	urldate = {2025-05-14},
	journal = {Physical Chemistry Chemical Physics},
	author = {Weigend, Florian and Ahlrichs, Reinhart},
	month = aug,
	year = {2005},
	pages = {3297--3305},
}

@article{marie_excited_2023,
	title = {Excited {States}, {Symmetry} {Breaking}, and {Unphysical} {Solutions} in {State}-{Specific} {CASSCF} {Theory}},
	volume = {127},
	issn = {1089-5639},
	url = {https://doi.org/10.1021/acs.jpca.3c00603},
	doi = {10.1021/acs.jpca.3c00603},
	number = {20},
	urldate = {2025-01-08},
	journal = {The Journal of Physical Chemistry A},
	author = {Marie, Antoine and Burton, Hugh G. A.},
	month = may,
	year = {2023},
	pages = {4538--4552},
}

@article{perego_data_2024,
	title = {Data efficient machine learning potentials for modeling catalytic reactivity via active learning and enhanced sampling},
	volume = {10},
	copyright = {2024 The Author(s)},
	issn = {2057-3960},
	url = {https://www.nature.com/articles/s41524-024-01481-6},
	doi = {10.1038/s41524-024-01481-6},
	language = {en},
	number = {1},
	urldate = {2025-01-07},
	journal = {npj Computational Materials},
	author = {Perego, Simone and Bonati, Luigi},
	month = dec,
	year = {2024},
	pages = {1--13},
}

@article{geng_intrinsic_2019,
	title = {Intrinsic {Reactivity} of {Diatomic} 3d {Transition}-{Metal} {Carbides} in the {Thermal} {Activation} of {Methane}: {Striking} {Electronic} {Structure} {Effects}},
	volume = {141},
	issn = {0002-7863},
	shorttitle = {Intrinsic {Reactivity} of {Diatomic} 3d {Transition}-{Metal} {Carbides} in the {Thermal} {Activation} of {Methane}},
	url = {https://doi.org/10.1021/jacs.8b11739},
	doi = {10.1021/jacs.8b11739},
	number = {1},
	urldate = {2025-01-07},
	journal = {Journal of the American Chemical Society},
	author = {Geng, Caiyun and Weiske, Thomas and Li, Jilai and Shaik, Sason and Schwarz, Helmut},
	month = jan,
	year = {2019},
	pages = {599--610},
}

@incollection{roos_complete_1987,
	title = {The {Complete} {Active} {Space} {Self}-{Consistent} {Field} {Method} and its {Applications} in {Electronic} {Structure} {Calculations}},
	copyright = {Copyright © 1987 by John Wiley \& Sons, Ltd.},
	isbn = {978-0-470-14294-3},
	url = {https://onlinelibrary.wiley.com/doi/abs/10.1002/9780470142943.ch7},
	language = {en},
	urldate = {2025-01-07},
	booktitle = {Advances in {Chemical} {Physics}},
	publisher = {John Wiley \& Sons, Ltd},
	author = {Roos, Björn O.},
	year = {1987},
	pages = {399--445},
}

@article{sharma_multiconfiguration_2021,
	title = {Multiconfiguration {Pair}-{Density} {Functional} {Theory}},
	volume = {72},
	issn = {0066-426X, 1545-1593},
	url = {https://www.annualreviews.org/content/journals/10.1146/annurev-physchem-090419-043839},
	doi = {10.1146/annurev-physchem-090419-043839},
	language = {en},
	number = {Volume 72, 2021},
	urldate = {2025-01-07},
	journal = {Annual Review of Physical Chemistry},
	author = {Sharma, Prachi and Bao, Jie J. and Truhlar, Donald G. and Gagliardi, Laura},
	month = apr,
	year = {2021},
	pages = {541--564},
}

@article{li_manni_multiconfiguration_2014,
	title = {Multiconfiguration {Pair}-{Density} {Functional} {Theory}},
	volume = {10},
	issn = {1549-9618},
	url = {https://doi.org/10.1021/ct500483t},
	doi = {10.1021/ct500483t},
	number = {9},
	urldate = {2025-01-07},
	journal = {Journal of Chemical Theory and Computation},
	author = {Li Manni, Giovanni and Carlson, Rebecca K. and Luo, Sijie and Ma, Dongxia and Olsen, Jeppe and Truhlar, Donald G. and Gagliardi, Laura},
	month = sep,
	year = {2014},
	pages = {3669--3680},
}

@article{saade_excited_2024,
	title = {Excited {State}-{Specific} {CASSCF} {Theory} for the {Torsion} of {Ethylene}},
	volume = {20},
	issn = {1549-9618},
	url = {https://doi.org/10.1021/acs.jctc.4c00212},
	doi = {10.1021/acs.jctc.4c00212},
	number = {12},
	urldate = {2025-01-07},
	journal = {Journal of Chemical Theory and Computation},
	author = {Saade, Sandra and Burton, Hugh G. A.},
	month = jun,
	year = {2024},
	pages = {5105--5114},
}

@article{roos_complete_1980,
	title = {A complete active space {SCF} method ({CASSCF}) using a density matrix formulated super-{CI} approach},
	volume = {48},
	issn = {0301-0104},
	url = {https://www.sciencedirect.com/science/article/pii/0301010480800450},
	doi = {10.1016/0301-0104(80)80045-0},
	number = {2},
	urldate = {2025-01-07},
	journal = {Chemical Physics},
	author = {Roos, Björn O. and Taylor, Peter R. and Sigbahn, Per E. M.},
	month = may,
	year = {1980},
	pages = {157--173},
}

@article{bofill_unrestricted_1989,
  author  = {Bofill, Josep M. and Pulay, Peter},
  title   = {The unrestricted natural orbital--complete active space (UNO--CAS) method: An inexpensive alternative to the complete active space--self-consistent-field (CAS--SCF) method},
  journal = {The Journal of Chemical Physics},
  year    = {1989},
  volume  = {90},
  number  = {7},
  pages   = {3637--3646},
  doi     = {10.1063/1.455822}
}

@article{abrams_natural_2004,
  author  = {Abrams, Micah L. and Sherrill, C. David},
  title   = {Natural orbitals as substitutes for optimized orbitals in complete active space wavefunctions},
  journal = {Chemical Physics Letters},
  year    = {2004},
  volume  = {395},
  number  = {4-6},
  pages   = {227--232},
  doi     = {10.1016/j.cplett.2004.07.081}
}

@article{jangid_linearized_2026,
  author  = {Jangid, Bhavnesh and Hennefarth, Matthew R. and Hermes, Matthew R. and Truhlar, Donald G. and Gagliardi, Laura},
  title   = {Linearized Pair-Density Functional Theory with Spin--Orbit Coupling},
  journal = {Journal of Chemical Theory and Computation},
  year    = {2026},
  volume  = {22},
  number  = {1},
  pages   = {318--333},
  doi     = {10.1021/acs.jctc.5c01633}
}

\end{document}


\maketitle

\tableofcontents
\clearpage

\section{Orbital Overlap Quantification}
\begin{figure}[htbp]
  \centering
  \includegraphics[width=\linewidth]{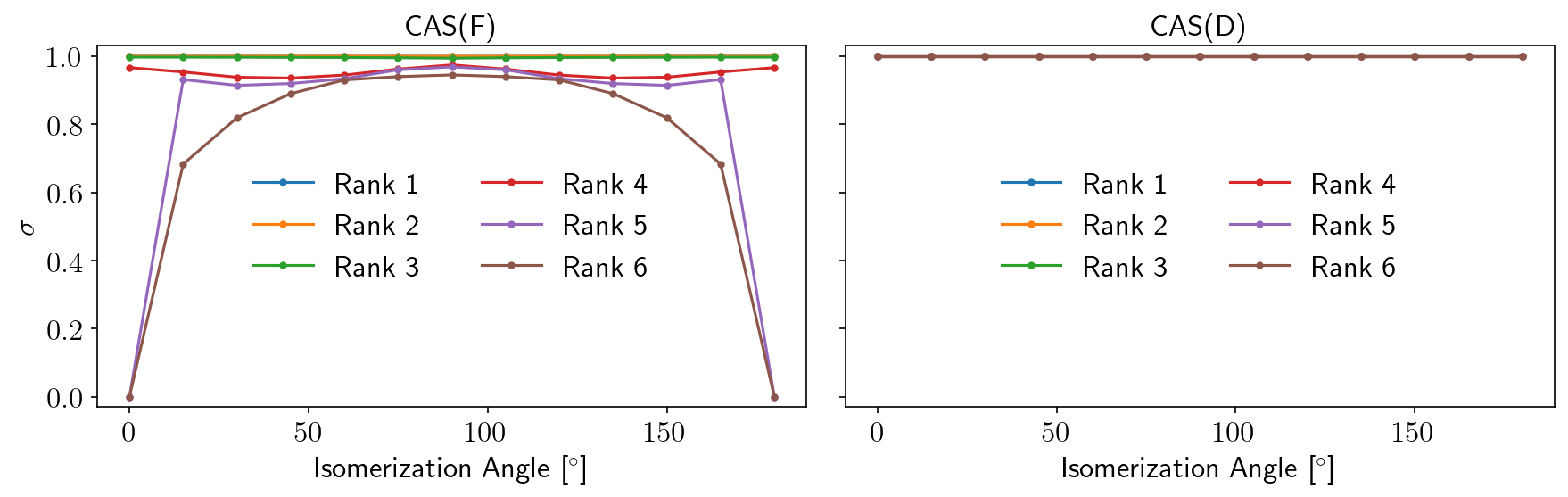}
  \caption{Active-space overlap singular values along the fulvene isomerization coordinate. The left and right panels compare active spaces reconstructed from the effective Fock matrix ($\mathbf{F}$) and the state-averaged 1-RDM ($\mathbf{D}$), respectively, with the reference SA-CASSCF active space. Each curve represents one of the six singular values, ranked independently at each geometry. Values near unity indicate close agreement with the reference active space.}
  \label{fig:fulvene-active-space-overlap}
\end{figure}

We quantify the agreement between the reference SA-CASSCF active space and the active spaces reconstructed from either the effective Fock matrix ($\mathbf{F}$) or the state-averaged one-particle reduced density matrix ($\mathbf{D}$). Let, $\mathbf{C}_{\mathrm{ref}}$ and $\mathbf{C}_{X}$ be the coefficients of the reference and reconstructed active orbitals, respectively, expressed in the atomic-orbital basis. The corresponding active-space overlap matrix is
\begin{equation}
  \mathbf{O}_{X}
  =
  \mathbf{C}_{\mathrm{ref}}^{\dagger}
  \mathbf{S}
  \mathbf{C}_{X},
\end{equation}
where $\mathbf{S}$ is the AO overlap matrix. To compare the two active subspaces independently of orbital rotations, we compute the singular-value decomposition
\begin{equation}
  \mathbf{O}_{X}
  =
  \mathbf{U}_{X}
  \boldsymbol{\Sigma}_{X}
  \mathbf{V}_{X}^{\dagger},
  \qquad
  \boldsymbol{\Sigma}_{X}
  =
  \operatorname{diag}(\sigma_{X,1},\ldots,\sigma_{X,6}).
\end{equation}
The singular values correspond to the cosines of the principal angles between the reconstructed and reference active spaces,
\begin{equation}
  \sigma_{X,i}=\cos\theta_{X,i}.
\end{equation}
Values close to unity therefore indicate strong agreement between the two subspaces, while smaller values indicate increasing deviation. Across the fulvene scan (Figure 1C in main text), the singular values obtained from $\mathbf{D}$ remain close to unity, indicating near-complete recovery of the reference SA-CASSCF active space (\Cref{fig:fulvene-active-space-overlap}). In contrast, the $\mathbf{F}$ reconstruction shows smaller singular values at several geometries, including two nearly vanishing values near $0^\circ$. These results motivate the use of $\mathbf{D}$ as the representation for CAS-WR.

\section{Supplementary Methods}

\subsection{Electronic Structure}
All electronic structure calculations were performed using the \texttt{PySCF} package \cite{sunPySCF2018,sunRecent2020,sun_python_2026-1} (version 2.12.1), compiled with \texttt{Libcint}\cite{SunLibcint2015} (version 6.1.3) and \texttt{Libxc}\cite{LehtolaRecent2018,MarquesLibxc2012} (version 7.0.0). The electronic energy convergence threshold was set to $10^{-7}$ a.u. No spatial symmetry constraints were imposed, and a numerical quadrature grid level of six was employed. Density fitting was used to accelerate the electronic structure calculations.

\subsubsection{Diels-Alder Reaction}
The Diels--Alder reaction between butadiene and ethylene was modeled using multi-configuration pair-density functional theory (MC-PDFT) with the translated PBE (tPBE) on-top functional.\cite{li_manni_multiconfiguration_2014} The underlying multireference wave function was computed using a CASSCF(6e,6o) active space with the def2-TZVP basis set.\cite{weigend_balanced_2005} This active space is identical to that used in our previous work,\cite{seal_computing_2025} where the corresponding active orbitals are shown.

\subsubsection{\ce{TiC+}-catalyzed Methane Activation}
The \ce{TiC+}-catalyzed \ce{C-H} activation of methane was modeled using MC-PDFT with the tPBE on-top functional.\cite{li_manni_multiconfiguration_2014} The underlying multireference wave functions were obtained using a SA(2)-CASSCF calculation with a (7e,9o) active space and def2-TZVP basis set.\cite{weigend_balanced_2005} This electronic structure protocol is identical to that employed in our previous study of this reaction,\cite{seal_weighted_2025} where the corresponding active orbitals are also reported.

\subsubsection{10-hydroxybenzo[h]quinoline (HBQ)}
The intramolecular excited-state proton transfer of 10-hydroxybenzo[h]quinoline (HBQ) was described using linearized pair-density functional theory (L-PDFT).\cite{hennefarth_linearized_2023,clifford_ultrafast_2026} Calculations employed a SA(2)-CASSCF reference with a (4e,4o) active space and the 6-31G** basis set,\cite{hehre_selfconsistent_1972,ditchfield_selfconsistent_1971,hariharan_influence_1973} followed by L-PDFT using the tPBE functional,\cite{li_manni_multiconfiguration_2014} as used in Clifford \textit{et al.}\cite{clifford_ultrafast_2026}

\subsubsection{\ce{(TMEDA)Ni^{II}(CH3)2}}
\begin{figure}[h!]
    \centering
    \includegraphics[width=\linewidth]{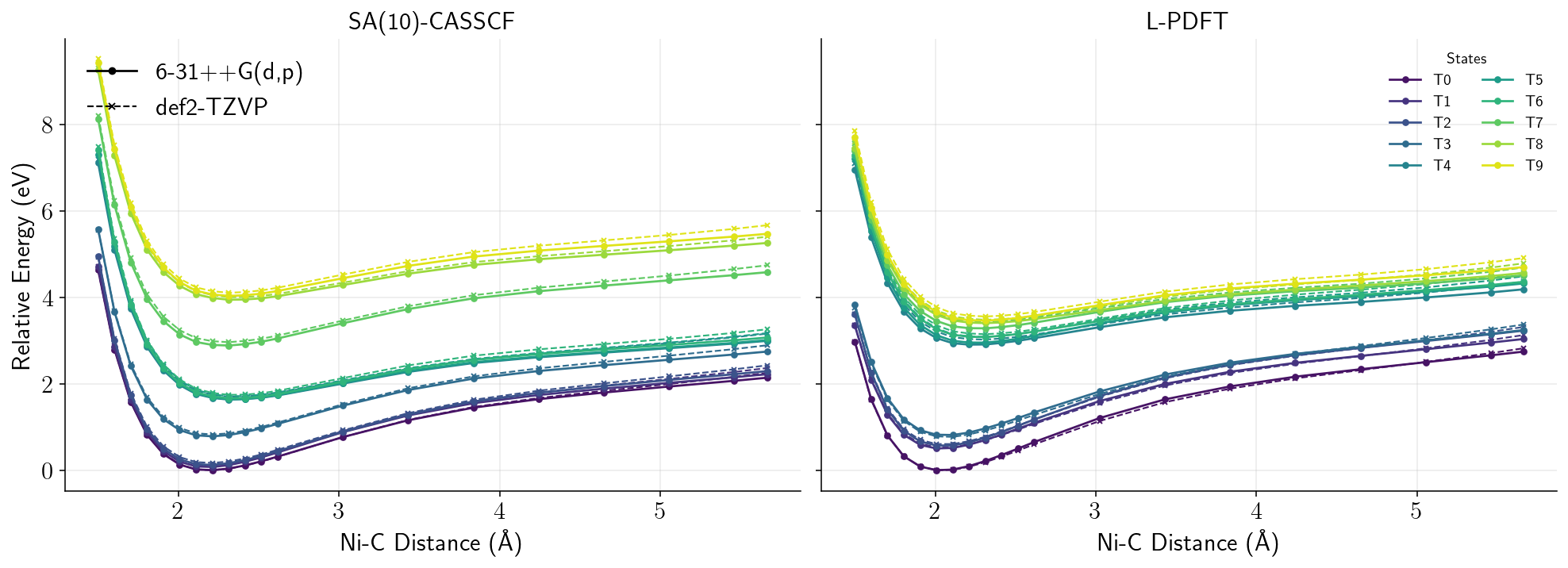}
    \caption{Comparison of the SA(10)-CASSCF and L-PDFT (with tPBE on-top functional) potential-energy surfaces along the \ce{Ni-C(CH3)} bond dissociation coordinate of the \ce{(TMEDA)Ni^{II}(CH3)2} complex, using the 6-31G++(d,p) and def2-TZVP basis sets.}
    \label{fig:def2tzvp}
\end{figure}

The electronic structure of \ce{(TMEDA)Ni^{II}(CH3)2} along Ni--C bond dissociation was described using L-PDFT with tPBE on-top functional.\cite{li_manni_multiconfiguration_2014} A state-averaged CASSCF reference over the ten lowest triplets was computed with (10e,6o) active space and 6-31G++(d,p) basis set.\cite{hehre_selfconsistent_1972,ditchfield_selfconsistent_1971,hariharan_influence_1973,clark_efficient_1983,rassolov_6-31g_1998} To assess the sensitivity of the potential-energy profile to the basis set, the Ni--C dissociation curve was additionally benchmarked using the def2-TZVP basis set\cite{weigend_balanced_2005} while retaining the same SA(10)-CAS(10e,6o) active-space specification (\Cref{fig:def2tzvp}).

\subsection{Kernel Ridge Regression}
Kernel ridge regression (KRR)\cite{murphy2012machine} was used to predict the underlying orbitals that describes the CAS wavefunction at a molecular geometry. For each training configuration, the external one-electron potential matrix, \(\mathrm{V}_{\mathrm{ext}}\), was used as the input representation, while the corresponding state-averaged one-particle reduced density matrix (1-RDM),
\begin{equation}
    \overline{\mathbf D}
    =
    \sum_I w_I \mathbf D_I,
\end{equation}
served as the regression target, where \(\mathbf D_I\) is the 1-RDM of electronic state \(I\) and \(w_I\) is its state-averaging weight, with \(\sum_I w_I = 1\). Both matrices were vectorized prior to training, such that the model learned the mapping
\begin{equation}
    \mathrm{vec}\!\left(\mathrm{V}_{\mathrm{ext}}\right)
    \longmapsto
    \mathrm{vec}\!\left(\overline{\mathbf D}\right).
\end{equation}

The KRR model was implemented using the \texttt{KernelRidge} estimator in \texttt{scikit-learn}\cite{pedregosa2011scikit} (version 1.7.2) with a radial basis function kernel. The kernel width and ridge regularization parameters were selected by cross-validation over $\alpha,\gamma \in \left\{10^{-6},10^{-5},10^{-4},10^{-3}\right\}$
using the negative mean-squared error as the scoring criterion. Following prediction, the vectorized output was reshaped into the 1-RDM and symmetrized. Natural orbitals were then obtained by diagonalizing the predicted density matrix. The resulting orbitals were used directly for the subsequent CASCI calculation. MC-PDFT or L-PDFT energies were then evaluated from the resulting CASCI wave functions, as appropriate. Thus, KRR predicts the state-averaged 1-RDM from which the CAS wave function is reconstructed.

\section{Active Learning Workflow and Statistics}

\subsection{1-RDM Uncertainty Guided Active Learning}

We construct the CAS-WR training set using a two-stage active-learning procedure that combines uncertainty in the predicted electronic structure with structural diversity. This strategy is inspired by the DEAL framework for selecting informative configurations in machine-learned interatomic potential training sets.\cite{seal_weighted_2025,perego_data_2024} Here, we adapt this approach to electronic-structure learning by using disagreement in the predicted one-particle reduced density matrix (1-RDM) to identify electronically uncertain configurations, followed by a structural-diversity criterion to remove redundant candidates.

In the first stage, a committee of $M$ independently trained models is constructed,
\begin{equation}
\left\{
\widehat{\mathbf D}^{(1)},
\widehat{\mathbf D}^{(2)},
\ldots,
\widehat{\mathbf D}^{(M)}
\right\}.
\end{equation}
For a geometry $\mathbf R_i$, the committee-averaged prediction is
\begin{equation}
\overline{\mathbf D}(\mathbf R_i)
=
\frac{1}{M}
\sum_{m=1}^{M}
\widehat{\mathbf D}^{(m)}(\mathbf R_i),
\end{equation}
and the committee disagreement is quantified as
\begin{equation}
\sigma_D(\mathbf R_i)
=
\frac{1}{M}
\sum_{m=1}^{M}
\left\|
\widehat{\mathbf D}^{(m)}(\mathbf R_i)
-
\overline{\mathbf D}(\mathbf R_i)
\right\|_{\mathrm F},
\end{equation}
where $\|\cdot\|_{\mathrm F}$ denotes the Frobenius norm. Configurations satisfying
\begin{equation}
\sigma_D(\mathbf R_i)
>
\alpha
\left\langle
\sigma_D
\right\rangle_{\mathrm{pool}}
\end{equation}
are retained as electronically uncertain candidates, where $\alpha$ is the selection threshold and $\langle \sigma_D\rangle_{\mathrm{pool}}$ denotes the mean committee disagreement over the candidate pool.

\begin{figure}[h!]
    \centering
    \includegraphics[width=0.9\linewidth]{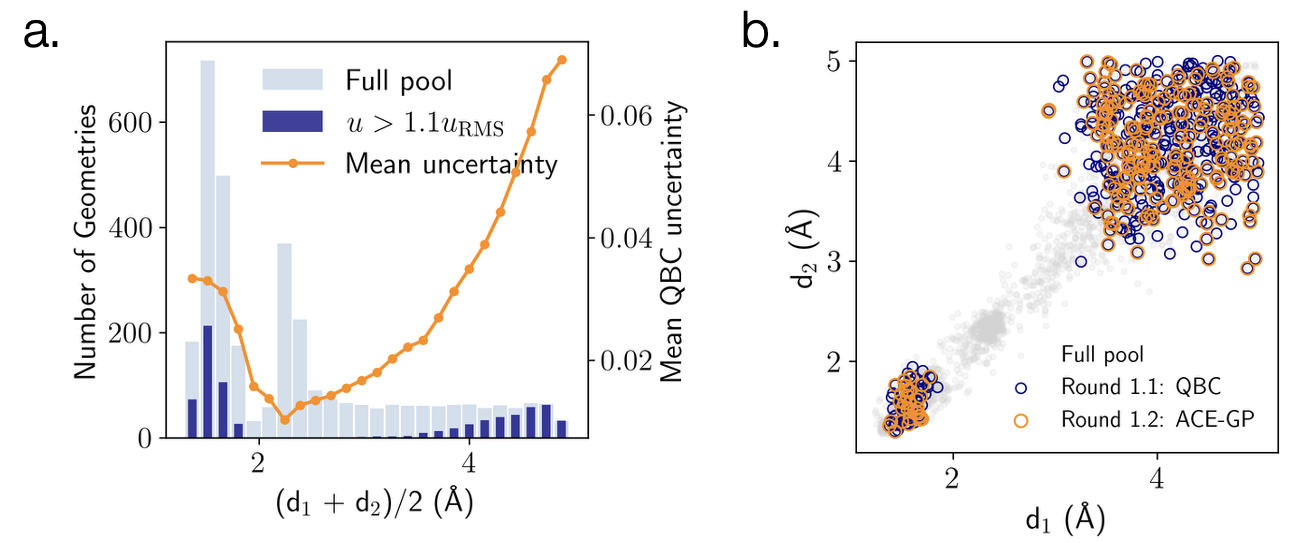}
    \caption{Representative active-learning round for the Diels--Alder reaction. (a) Mean query-by-committee (QBC) uncertainty as a function of ($\mathrm{d}_1$+$\mathrm{d}_2$)/2. (b) Geometries selected at the two stages, shown in the space of the two forming C--C distances, $\mathrm{d}_1$ and $\mathrm{d}_2$.}
    \label{fig:deal}
\end{figure}

In the second stage, structural redundancy within the uncertain candidate set is reduced using a descriptor-space diversity criterion. Each geometry is mapped to the ACE representation $\mathbf z_i$, and the Gaussian-process posterior variance is evaluated as
\begin{equation}
\sigma_{\mathrm{GP}}^2(\mathbf z)
=
k(\mathbf z,\mathbf z)
-
\mathbf k_{\mathbf zS}
\mathbf K_{SS}^{-1}
\mathbf k_{S\mathbf z},
\end{equation}
where $S$ denotes the set of previously selected configurations. Candidates with large posterior variance are added iteratively, preferentially selecting configurations from regions of descriptor space that are poorly represented in the existing training set. A representative active-learning round is shown in \Cref{fig:deal}, illustrating the uncertainty-based candidate identification and subsequent diversity-based selection.

For the Diels--Alder system, the initial training set (Round 0) was generated from NVE molecular dynamics trajectories initiated from the transition-state geometry. Initial velocities were sampled from a Maxwell--Boltzmann distribution at 300 K, and the dynamics were propagated using MC-PDFT analytical gradients with a 0.5 fs time step. Geometries were retained only when both forming C--C distances, were less than 5~\AA; the same geometric filter was applied in all subsequent active-learning rounds. Candidate pools for later rounds were constructed from a combination of picosecond-scale molecular dynamics trajectories initiated from configurations identified as uncertain in the preceding round, with simulations performed at 300 and 600 K; normal-mode-like perturbations about the transition-state geometry; and high-energy randomly perturbed structures generated using GFN2-xTB.\cite{Bannwarth2019GFN2xTB} Configurations selected by the active-learning procedure were then labeled with CASSCF/MC-PDFT calculations, where each calculation was initialized using checkpoint orbitals from the nearest labeled geometry in the preceding round.

\subsection{CAS-WR Active Learning Statistics}

\begin{figure}[h!]
    \centering
    \includegraphics[width=0.9\linewidth]{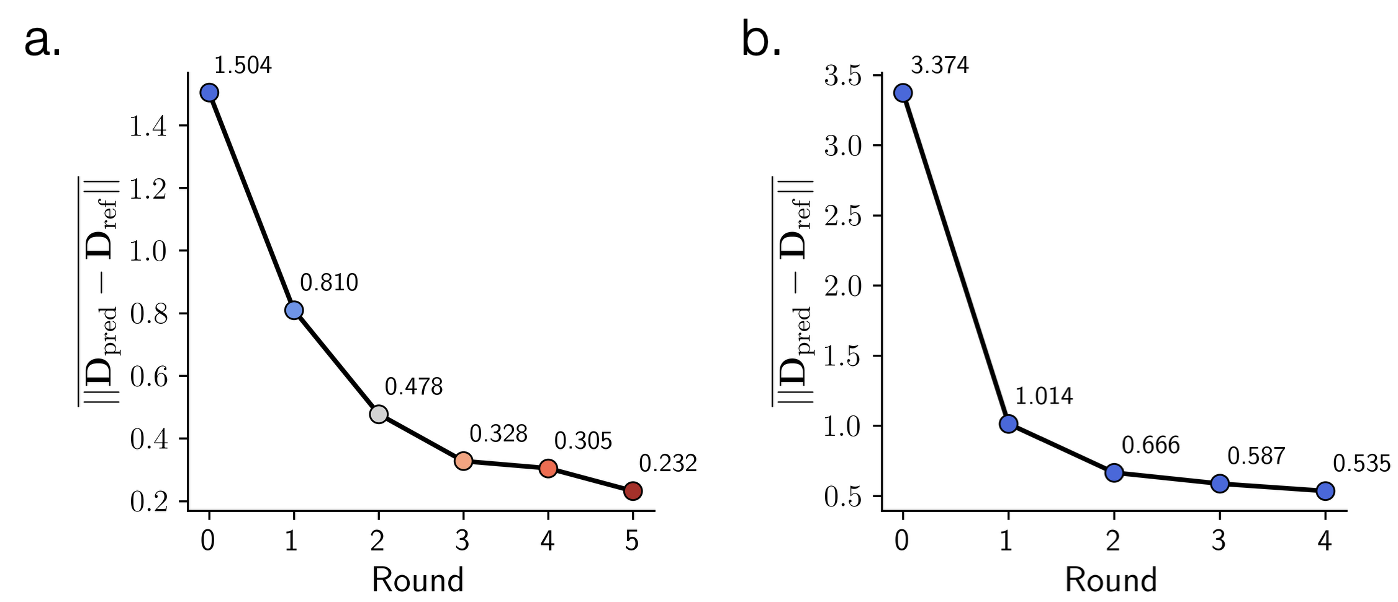}
    \caption{Evolution of the held-out state-averaged 1-RDM Frobenius-norm error during active learning. (a) \ce{TiC+ + CH4} reaction dataset at SA(2)-CAS(7e,9o)/def2-TZVP, using data from Seal \textit{et al.}\cite{seal_weighted_2025} (b) Diels--Alder reaction dataset obtained from the 1-RDM uncertainty-based active-learning procedure described in this work, using CAS(6e,6o)/def2-TZVP.}
    \label{fig:rdm_error}
\end{figure}

The performance of CAS-WR at each active-learning round was evaluated on a held-out test set using two metrics. First, the accuracy of the predicted electronic structure was quantified directly from the state-averaged 1-RDM using the mean Frobenius-norm error,
\begin{equation}
    \epsilon_D
    =
    \frac{1}{N_{\mathrm{test}}}
    \sum_{i=1}^{N_{\mathrm{test}}}
    \left\|
    \overline{\mathbf D}^{\,\mathrm{pred}}_i
    -
    \overline{\mathbf D}^{\,\mathrm{ref}}_i
    \right\|_{\mathrm F},
\end{equation}
where $\overline{\mathbf D}^{\,\mathrm{pred}}_i$ and $\overline{\mathbf D}^{\,\mathrm{ref}}_i$ denote the predicted and reference state-averaged 1-RDMs, respectively. The evolution of the held-out 1-RDM Frobenius-norm error over successive active-learning rounds is shown in \Cref{fig:rdm_error} for both the \ce{TiC+ + CH4} and Diels--Alder reaction.

Second, to assess how errors in the predicted 1-RDM propagate to downstream tasks post-CI optimization, we evaluated the mean absolute error (MAE) in the energies obtained after reconstructing the orbital space and performing the CI optimization and MC-PDFT/L-PDFT calculations (reported in main text),
\begin{equation}
    \mathrm{MAE}_{E}
    =
    \frac{1}{N_{\mathrm{test}}}
    \sum_{i=1}^{N_{\mathrm{test}}}
    \left|
    E_i^{\mathrm{CAS\text{-}WRM}}
    -
    E_i^{\mathrm{ref}}
    \right|.
\end{equation}

\section{Supplementary Figures}

\begin{figure}[h!]
    \centering
    \includegraphics[width=0.8\linewidth]{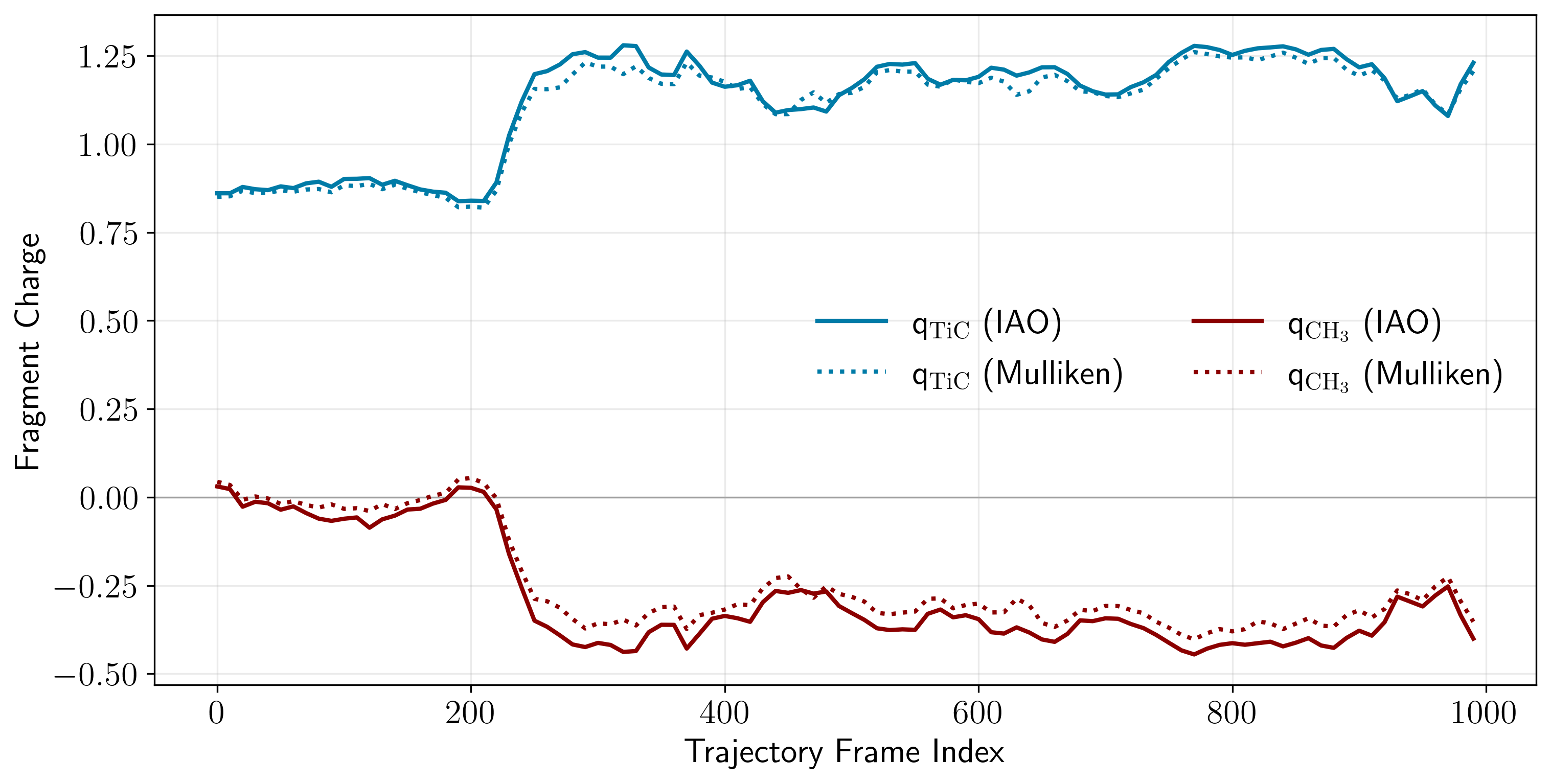}
    \caption{Comparison of fragment charges obtained using intrinsic atomic orbital (IAO) and Mulliken population analyses (reported in main text) along a representative reactive trajectory for \ce{TiC+ + CH4}. The charges of the \ce{TiC} and \ce{CH3} fragments are shown as a function of simulation timestep computed at the CAS(7e,9o)/def2-TZVP level of theory.}
    \label{fig:placeholder}
\end{figure}

\begin{figure}[h!]
    \centering
    \includegraphics[width=0.6\linewidth]{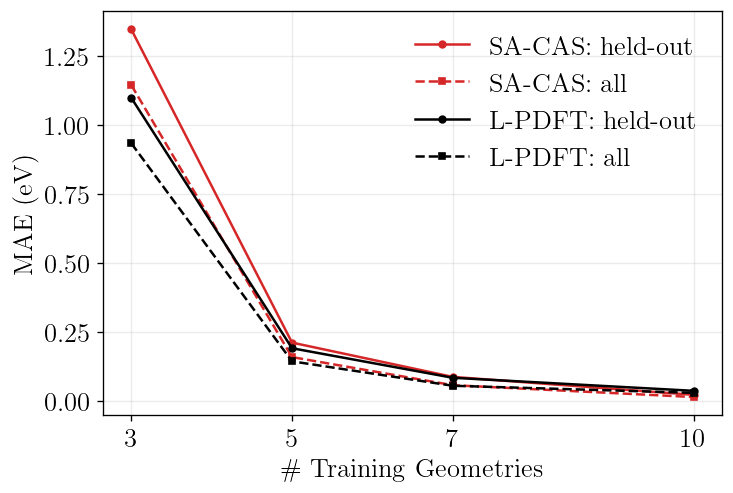}
    \caption{Mean absolute error (MAE) in the SA-CASCI and L-PDFT energies for \ce{(TMEDA)Ni^{II}(CH3)2} as a function of the number of CAS-WR training geometries. The initial training set contains the two endpoint geometries along the \ce{Ni-C(CH3)} bond-dissociation coordinate, with additional geometries introduced successively from within the dissociation interval to give training sets of 3, 5, 7, and 10 geometries. At each stage, the MAE is evaluated over the remaining geometries not included in the training set (held-out) and over the full dissociation profile (all).}
    \label{fig:placeholder}
\end{figure}

\newpage
\bibliography{ref}